\documentclass[preprint,12pt]{elsarticle}
\pdfoutput=1 

\usepackage{soul}
\usepackage[normalem]{ulem}
\usepackage[T1]{fontenc}
\usepackage[utf8]{inputenc}
\usepackage{multirow}
\usepackage[english]{babel}
\usepackage{hyperref}
\usepackage{amsmath}
\usepackage{amssymb}
\usepackage{epsfig}
\usepackage{ccaption}
\usepackage{lscape}
\usepackage{longtable}
\usepackage{float}
\usepackage{graphics,psfrag,rotating}
\usepackage{graphicx}
\usepackage{dcolumn}
\usepackage{bm}
\usepackage{siunitx}
\usepackage{array,multirow}
\usepackage{xspace}
\usepackage[dvipsnames]{xcolor}
\usepackage{placeins}
\usepackage{afterpage}
\usepackage{etoolbox}
\usepackage{colortbl}
\usepackage{array,longtable}
\definecolor{Gray}{gray}{0.85}

\begin{document}


\title{Uncertainties in the Galactic Dark Matter Distribution: an Update}

\author[a]{Mar\'ia~Benito}
\ead{mariabenitocst@gmail.com}
\author[b]{Fabio~Iocco}
\ead{fabio.iocco@unina.it}
\author[c]{Alessandro~Cuoco}
\ead{alessandro.cuoco@unito.it}

\affiliation[a]{organization={National Institute of Chemical Physics and Biophysics},
             addressline={R\"avala 10},
             city={Tallinn},
             postcode={10143},
             country={Estonia}}
             
\affiliation[b]{organization={Universit\`a di Napoli ``Federico II'' \& INFN Sezione di Napoli},
             addressline={Complesso Universitario di Monte S.~Angelo, via Cintia,},
             city={Napoli},
             postcode={80126},
             country={Italy}}
             
\affiliation[c]{organization={Dipartimento di Fisica, Universit\`a di Torino, \& INFN Sezione di Torino},
             addressline={Via P. Giuria 1},
             city={Torino},
             postcode={10125},
             country={Italy}}

\begin{abstract}
We present here a quantitative estimate of the impact of uncertainties of astrophysical nature on the determination of the dark matter distribution within our Galaxy, the Milky Way. 
Based on an update of a previous analysis, this work is motivated by recent new determinations of astrophysical quantities of relevance --such as the Galactic parameters (R$_0$,V$_0$)-- from the GRAVITY collaboration and the GAIA satellite, respectively. We find that even with these state--of--the--art determination and a range of uncertainties --both statistical and systematic-- much narrowed with respect to previous literature, the uncertainties on the dark matter distribution and their impact on searches of physics beyond the standard model stays sizable.    
\end{abstract}

\maketitle

\section{Introduction}
\label{sec:intro}
\par The determination of the gravitational structure of our host Galaxy, the Milky Way (MW), is a very interesting endeavor by itself, and at the same time it has implications that reverberate from Cosmology to Particle Physics. 
The gravitational potential of the MW can not be explained by the presence of stars and gas alone, beyond the innermost $\sim$ 5 kpc, \cite{2015NatPh..11..245I}.
This is generally imputed to the presence of a component of unknown nature dubbed {\it Dark} Matter (DM). On the one hand, this component of matter cannot be accommodated within the Standard Model of Particle Physics. This has motivated direct and indirect particle searches, together with collider experiments, that aim to understand its nature. 
Synergies between these efforts have constrained the parameter space of several extensions of the Standard Model. However, these attempts are hampered since the interpretation of data from direct and indirect searches depends on the distribution of DM in the Galaxy.
On the other hand, the distribution of DM in galaxies is a prediction of the $\Lambda$CDM model, thus it provides an important test of consistency of the cosmological framework.

From the above, it proceeds that the distribution of the DM within our Galaxy is of relevance, beside its intrinsic value ``{\it per se}'', as an ancillary quantity for other fields.
Determined through techniques that rely on astrophysical observations \cite{2015ApJ...803L...3P, 2013ApJ...779..115B, 2015JCAP...12..001P, 2017PDU....15...90I, 2017JCAP...02..007B, 2019arXiv191204296K, 2019JCAP...09..046K, Nesti:2013uwa, Silverwood:2015hxa, Catena:2009mf, 2018MNRAS.478.1677S, 2019JCAP...10..037D, 2020MNRAS.tmp.1873W, 2020arXiv201211477D}, 
the DM distribution is unavoidably affected by the uncertainties that plague such observations.
Such uncertainties do propagate into other quantities that rely on the use of the DM distribution, and hence the original ignorance on astrophysical quantities propagate to quantities of seemingly unrelated nature, such as the so-called DM $J$-factor which regulates the amount of DM annihilation signal and thus the expected yield of e.g. $\gamma$-ray photons, neutrinos or antiprotons, which would reveal the presence of DM itself, or the local DM density $\rho_0$ which, instead, dictates the expected number of events in underground direct detection experiments \cite{2019JCAP...03..033B}.

\par The principles of the above are very well known, yet a specific quantitative approach, systematically estimating the effect of all the observables into play, is not thoroughly adopted.
In a previous work \cite{2017JCAP...02..007B} we had proposed a first quantitative estimate  of the impact of astrophysical uncertainties on specific scenarios for the DM nature. Later, we had proposed a systematic approach to the astrophysical quantities in play in the empirical determination of the DM distribution \cite{2019JCAP...03..033B} (hereafter Paper I). In Paper I, we also presented a likelihood function that can be used in the particle interpretation of data coming from direct and indirect searches in order to self-consistently include astrophysical uncertainties that affect our determination of the DM distribution \cite{2019JCAP...04..040F, 2019PhRvD.100d3020A, 2019PhRvD.100j3014L, 2019arXiv191209486A, 2020PhRvD.101d3526S, 2020arXiv200306614A, 2020arXiv200508824B, 2020arXiv200606836G, 2019PhRvL.122q1801B}.

In this new paper, we present an approach very similar to that of Paper I, slightly modified from the technical point of view, and including the recent most determinations of some of the astrophysical quantities that have a bigger impact in the determination of the DM distribution, namely the Sun's distance to the Galactic center, $R_0$, and its circular velocity $V_0$. We anticipate that despite uncertainties on these quantities are narrowed, the remaining uncertainties on the DM distribution are sizable, thus still affecting searches for its nature. 

\par This paper is structured as follows:
in section \ref{sec:meth} we describe the new methodology;
in section \ref{sec:data} we present the new observations we adopt for this determination;
in section \ref{sec:results} we present our results, also comparing the state--of--the--art and the improvement of knowledge with respect to previous determinations. 
We present our conclusions in section \ref{ref:conclusions}, while in Appendix \ref{App:burkert_profile} and \ref{App:einasto_profile} we discuss the case of alternative DM profiles,  and in Appendix \ref{App:bayesian}  we provide  the results of a Bayesian analysis and compare them with the frequentist analysis.

\section{Methodology}
\label{sec:meth}

We closely follow the data-driven analysis presented in Paper I in order to quantify astrophysical uncertainties on our determination of the DM distribution in the MW. In particular, constraints on the distribution of DM are obtained with the well-known rotation curve (RC) method, by comparing the observed RC of the MW with predicted velocities expected to be caused by the baryonic and DM components of the Galaxy.

We adopt the data from the {\tt galkin} compilation \cite{2015NatPh..11..245I,2017SoftX...6...54P} for the observed RC. Observed velocities depend on the Sun's galactocentric distance $R_0$, its circular velocity $V_0$ and its peculiar motion $(U, V, W)_{\odot}$. 
The Sun's peculiar motion in the tangential direction $V_{\odot}$, $R_0$ and $V_0$ are related to the Sun's total angular velocity, $\Omega_{\rm 0, tot}$, by
\begin{equation}
    \Omega_{\rm 0, tot} = \frac{V_{\rm 0, tot}}{R_0} = \frac{V_0 + V_{\odot}}{R_0}.
    \label{eq:sun_angular_velocity}
\end{equation}
In Paper I \cite{2019JCAP...03..033B}, we fixed $\Omega_{\rm 0, tot}$, whose  value is known with a small uncertainty, and $V_{\odot}$, and we varied $R_0$ in the range 7.5-8.5 kpc. Each time $R_0$ is specified, $V_0$ was derived following the above equation. The generous range of variation for $R_0$ was in part compensating for   having kept fixed $V_{\odot}$, thus neglecting its uncertainty. In this work we rather fix $\Omega_{\rm 0, tot}$ and $R_0$, which has been recently precisely measured (see below), and we vary $V_0$ within measured uncertainties. Each time $V_0$ is specified, $V_{\odot}$ is obtained by means of equation \eqref{eq:sun_angular_velocity}.  That is, the quantities $V_0$ and $V_\odot$ self-consistently satisfy constraints on the Solar total velocity in the tangential direction, which is estimated with high precision \cite{DrimmelPoggio18}. 

We assume the DM is a smooth, spherically-symmetric component whose distribution is described by a generalized Navarro-Frenk-White (gNFW) profile \cite{1996MNRAS.278..488Z, 2001ApJ...555..504W} (in appendices \ref{App:burkert_profile} and \ref{App:einasto_profile} we show the results for the Burkert \cite{Burkert_1995} and Einasto \cite{1965TrAlm...5...87E} DM density profiles, respectively). For the baryonic matter, we adopt 
a set of several baryonic morphologies $\mathcal{M}_i$ -- motivated by observations -- that bracket the systematic uncertainty on the distribution of the baryonic mass in our Galaxy \cite{2015NatPh..11..245I}. 
A complete description of the the baryonic morphology catalog can be found in \cite{2015NatPh..11..245I, 2019JCAP...03..033B} and references therein.
We also account for the uncertainty on the total baryonic mass by normalizing the stellar disk profile to the stellar surface density at the Sun’s position $\Sigma_*$ and by normalizing the bulge mass using the microlensing optical depth towards the galactic center $\langle\tau\rangle$. 

Our analysis has, thus, the following free parameters: $V_0$, $\Theta$, $\mathcal{M}_i$, $\langle\tau\rangle$ and $\Sigma_*$; where $\Theta=(R_s, \rho_0, \gamma)$ correspond to the parameters of the DM density profile, i.e. the profile scale radius, the local DM density and the profile inner slope, respectively.
We scan a discrete grid composed of 50 values for $\rho_0$ linearly spaced in the range [0, 1] $\rm GeV/cm^3$, 50 values for $R_s$ logarithmically spaced in the range [5, 100] kpc, 15 values of $\gamma$ linearly spaced in the range [0, 1.5], 10 values of $V_0$ linearly spaced in the range [218, 240] km/s, and 30 morphologies $\mathcal{M}_i$. For $\langle\tau\rangle$ and $\Sigma_*$ we use 10 values each, linearly spaces in the range [-2$\sigma$, +2$\sigma$].
At each point of this seven-dimensional grid, observed and predicted rotation velocities are compared by means of a $\chi^2$ statistics given by
\begin{equation}
\begin{split}
    \chi^2_{\rm RC}(V_0, \Theta, \mathcal{M}_i, \langle\tau\rangle, \Sigma_*) = &\sum_j\frac{\left(\bar{w}_j(\Theta, \mathcal{M}_i, \langle\tau\rangle, \Sigma_*) - \bar{w}^{obs}_j(V_0)\right)^2}{\sigma_{\bar{w}_j}^2} \\
    &+ \frac{\left(\langle\tau\rangle - \langle\tau\rangle^{obs}\right)^2}{\sigma_{\langle\tau\rangle}^2} + \frac{\left(\Sigma_* - \Sigma_*^{obs}\right)^2}{\sigma_{\Sigma_*}^2},
\end{split}
\end{equation}
where $\bar{w}^{obs}_j$ is the measured angular velocity, with its corresponding uncertainty $\sigma_{\bar{w}_j}$, for a given radial RC bin $j$. 
For details on how the binned quantities $\bar{w}^{obs}_j$ and $\sigma_{\bar{w}_j}$ are derived from the {\tt galkin} compilation of measurements we refer the reader to Paper I. We also notice that for each different values of $V_0$ on the grid  the experimental angular velocities also change accordingly. We self-consistently take this effect  into account. Again, this is discussed in detail in Paper I. 
We adopt the values of the microlensing optical depth measurement provided in \cite{2005ApJ...631..879P}, i.e. $\langle\tau\rangle^{obs}=2.17^{+0.47}_{-0.38}\times 10^{-6}$, as well as the stellar surface density at the Sun's position provided by \cite{2013ApJ...779..115B}, namely $\Sigma_*^{obs}=(38\pm4){\rm M_{\odot}/pc^2}$. For simplicity, we symmetrize the error in the microlensing optical depth and adopt a standard deviation of $\sigma_{\langle\tau\rangle}=0.47$.

We employ a frequentist formalism and derive profile likelihoods.
For a thorough description of the statistical framework, we refer the interested reader to section 3 in Paper I.
Nonetheless, for completeness, we also present bayesian results, which do not rely on a grid but make use of  Monte Carlo scan techniques. The results of the Bayesian analysis are reported in appendix \ref{App:bayesian}.

\section{New observations}
\label{sec:data}

In this work we adopt the following new estimates of the relevant astrophysical quantities $(R_0, V_0)$:
\begin{itemize}
    \item[--] the Sun's galactocentric distance estimation obtained by the GRAVITY collaboration by measuring the Keplerian orbit of the S2 star in the innermost parsecs of the Galaxy \cite{GRAVITY2019}: 
    
    \begin{equation}
    R_0=8.178\pm0.013{\rm (stat)}\pm0.022{\rm (syst)}.
    \end{equation}
    \item[--] The Sun's circular velocity determined by means of a Jeans analysis that combines Gaia~\cite{2016A&A...595A...1G}, WISE~\cite{2010AJ....140.1868W} and 2MASS~\cite{2006AJ....131.1163S} photometry with spectral data from APOGEE~\cite{2017AJ....154...94M} for $\sim 23000$ red-giant stars with galactocentric distances between 5 and 25 kpc \cite{Eilers+19}: 
    \begin{equation}
    V_0=229.0\pm0.2\,{\rm km/s},
    \end{equation}
    with a systematic uncertainty in the range $2-5$\%.
\end{itemize}

We fix $R_0$ to the GRAVITY estimate\footnote{If we rather fix $R_0$ to the updated estimate given in \citep{GRAVITY2020} (i.e. $R_0=8.249\pm0.009{\rm (stat)}\pm0.045{\rm (syst)}$)~\citep{GRAVITY2020}, uncertainties in $\rho_0$ vary by less than $3\%$.} and we vary $V_0$ within measured uncertainties. 
We adopt as fiducial the $V_0$ range $[218, 240]\,{\rm km/s}$, chosen to encompass the conservative 5\% systematic uncertainty quoted in \cite{Eilers+19}, and incidentally coinciding with the range of 
values found in the literature (e.g. \cite{2012ApJ...759..131B, 2014ApJ...783..130R, BlandHawthornGerhard16, Schonrich2012, 2011AstL...37..254S, 2010ApJ...712..260K}).\footnote{Each time $V_0$ is specified, $V_{\odot}$ is derived -- according to equation~\eqref{eq:sun_angular_velocity} -- in order to satisfy constraints on the Solar total velocity. In particular, by varying $V_0$ in the range $[218, 240]\,\rm km/s$, $V_{\odot}$ varies in the range $[7, 29]\,\rm km/s$, which indeed perfectly brackets estimates from the literature (e.g. \cite{2005ApJ...629..268H, Schooenrich+2010, 2015ApJ...809..145T, 2015MNRAS.449..162H, 2014ApJ...793...51S, 2014ApJ...783..130R, 2015ApJ...800...83B})}

\begin{figure}[t]
\centering
\includegraphics[scale=0.45]{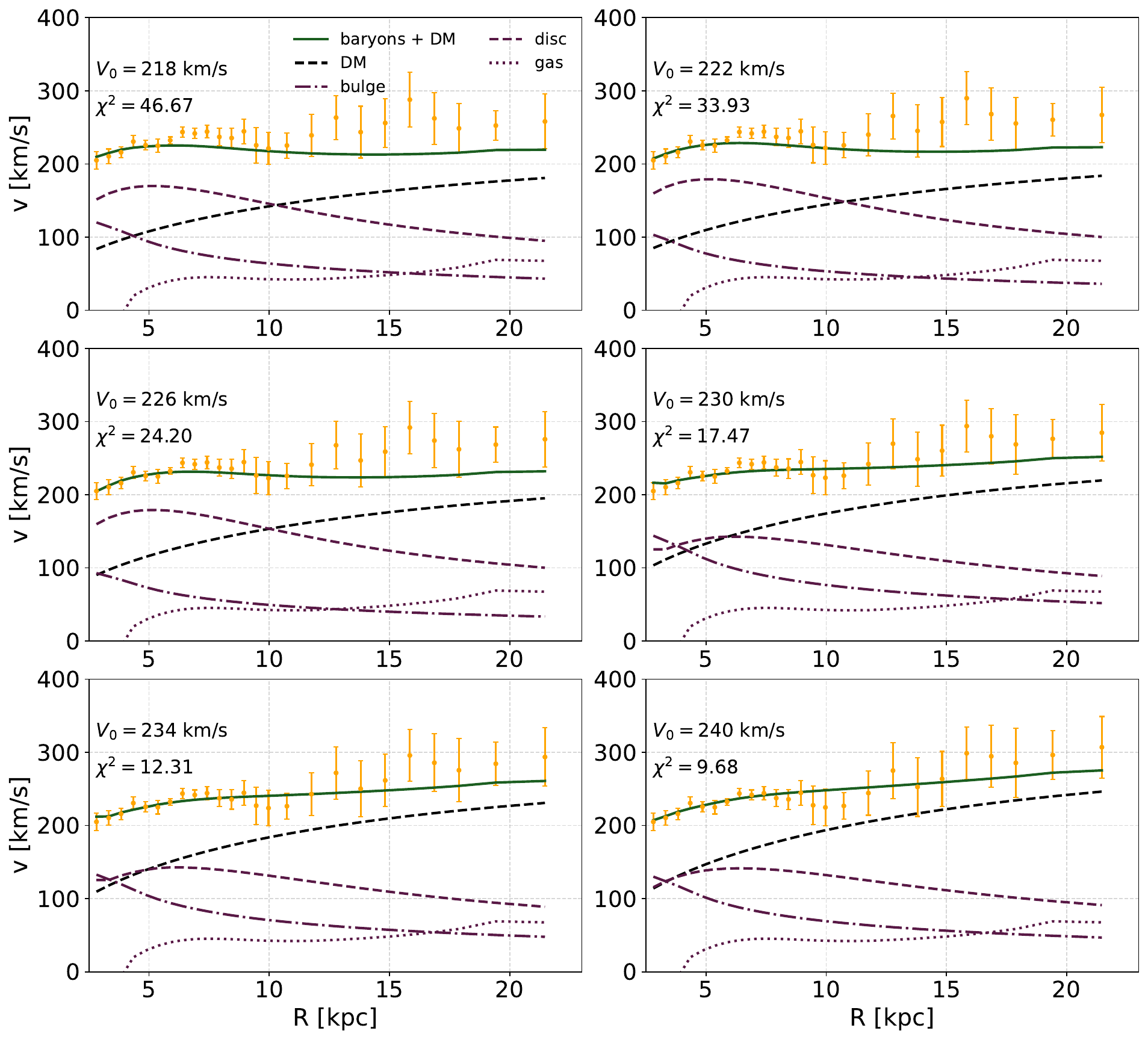}
\caption{Observed Rotation Curve and best-fit contributions of the bulge, disc, gas, DM, individually as well as summed together, for different values of $V_0$. The slope of the DM profile (gNFW) has been fixed to $\gamma=1$ for these plots (see text for more details).
}
\label{fig:RCfit}
\end{figure}

As in Paper I, the Solar total angular velocity is fixed to the precise result $\Omega_{\rm 0, tot}=30.24\pm0.12\,{\rm km/s/kpc}$, which is obtained by measuring the proper motion of Sgr A$^*$~\cite{ReidBrunthaler2004}.
The Sun's peculiar motion in the radial and vertical directions are fixed to $U_{\odot}=11.10\, \rm km/s$ and $W_{\odot}=7.25\, \rm km/s$ \cite{Schonrich2012}, respectively. These two quantities are measured with $\sim 10\%$ precision see e.g. \cite{BlandHawthornGerhard16} and references therein. By varying them within measured uncertainties, our results remain unaffected. It is to be noticed that whether $U_{\odot}$ and $W_{\odot}$ are measured with the indicated precision, large scatter surrounds the estimates of $V_\odot$. In fact, the range of $V_\odot$ values adopted in this work, which spans $\SI{22}{km/s}$, encompasses global and local estimates found in the literature \cite{2005ApJ...629..268H, Schooenrich+2010, 2015ApJ...809..145T, 2015MNRAS.449..162H, 2014ApJ...793...51S, 2014ApJ...783..130R, 2015ApJ...800...83B}. Sizeable uncertainties on this parameter might be explained by streaming motion induced by local substructures or/and spiral arms~\cite{BlandHawthornGerhard16}.

\begin{figure}[t]
\centering
\includegraphics[scale=0.35]{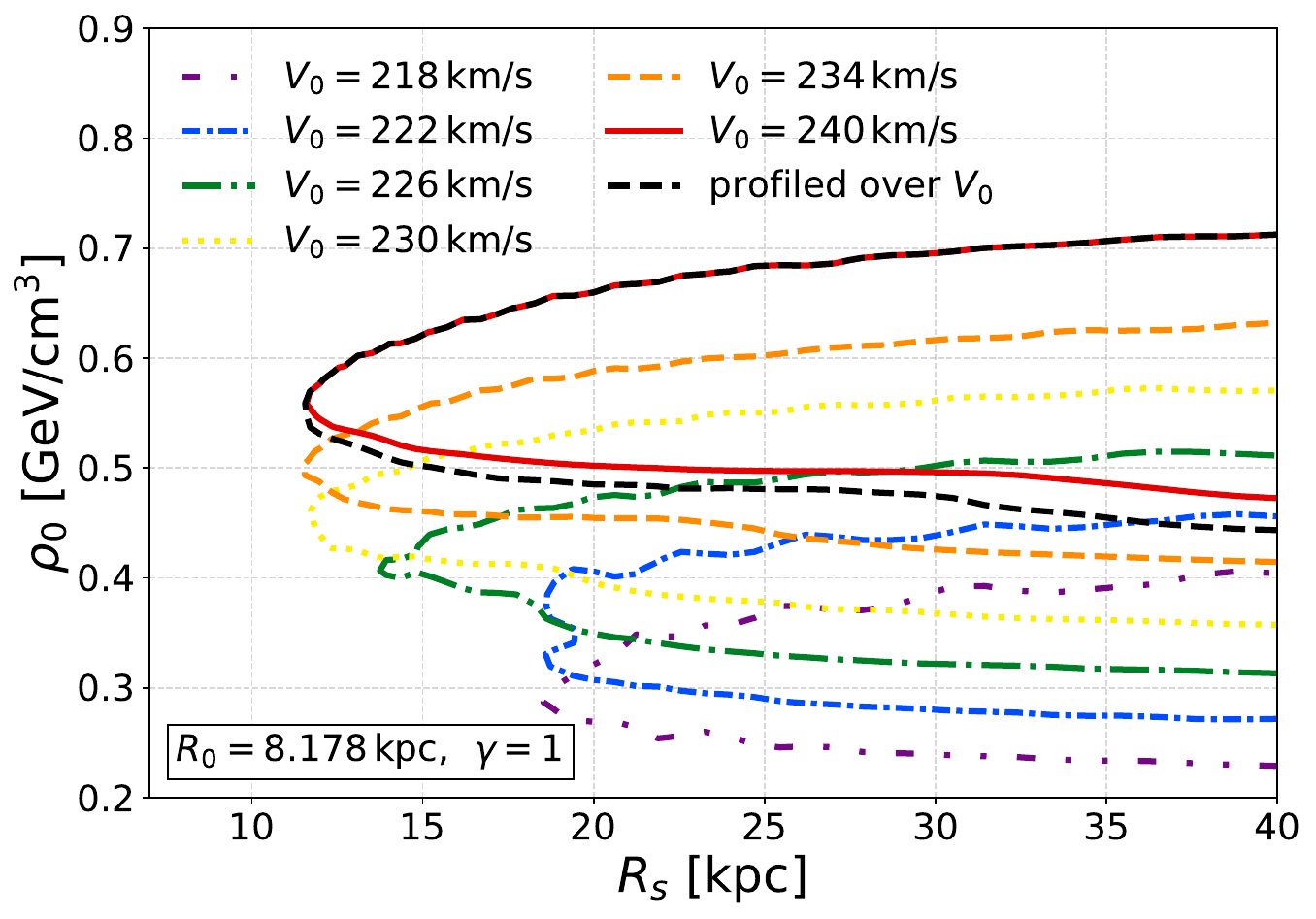}
\includegraphics[scale=0.35]{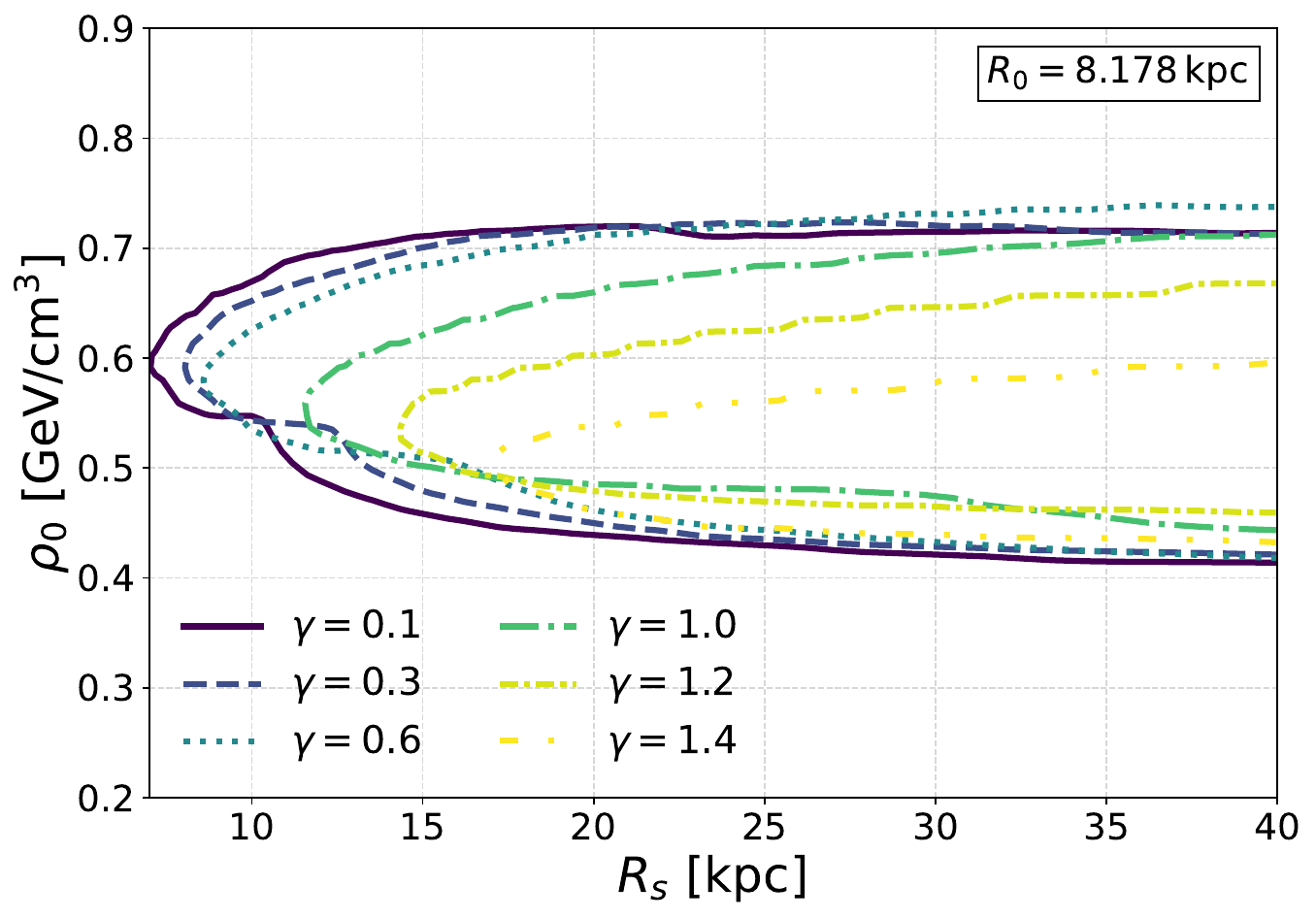}
\caption{Top panel: 2$\sigma$ contours in the $(R_s,\rho_0)$ plane for fixed $\gamma=1$ and $R_0={8.178}\,{\rm kpc}$, and for  various values of $V_0$, and profiled over $\mathcal{M}_i$, $\langle\tau\rangle$ and $\Sigma_*$. The dashed black line is the 2$\sigma$ contour further profiled over $V_0$. 
Bottom panel: 2$\sigma$ contours in the $(R_s,\rho_0)$ plane for fixed $R_0={8.178}\,{\rm kpc}$, for various values of $\gamma$, and profiled over $V_0$, $\mathcal{M}_i$, $\langle\tau\rangle$ and $\Sigma_*$. 
}
\label{fig:gNFWcontours}
\end{figure}

\section{Results}
\label{sec:results}

\par 
In this section we present our results. In figure \ref{fig:RCfit} we show some example of how the best fit RC compares with the observations for various fixed values of $V_0$.
The quality of the best-fit is good with a value of the $\chi^2$ of about 9 given the 25 data points.
In the top panel of figure~\ref{fig:gNFWcontours}, we show $2\sigma$ contours of the profile $\chi^2_{\rm RC}$ for fixed $\gamma=1$ and different $V_0$ values, i.e. 
\begin{equation*}
\chi^2_{\rm RC, prof}(V_0, R_s, \rho_0, \gamma=1),
\end{equation*}
where the remaining parameters $\mathcal{M}_i, \langle\tau\rangle, \Sigma_*$ have been profiled away, i.e., 
for given $V_0, R_s, \rho_s$ and $\gamma=1$, $\chi^2_{\rm RC}$ is minimized over $\mathcal{M}_i, \langle\tau\rangle, \Sigma_*$ to give  $\chi^2_{\rm RC, prof}$.
 We generalize our results for different $\gamma$ in the bottom panel of this same figure, where we show the $2\sigma$ contours of the $\chi^2_{\rm RC, prof}$ further profiled over $V_0$ for various values of $\gamma$.

\subsection{Comparison with Paper I}

\begin{figure}[t]
\centering
\includegraphics[scale=0.35]{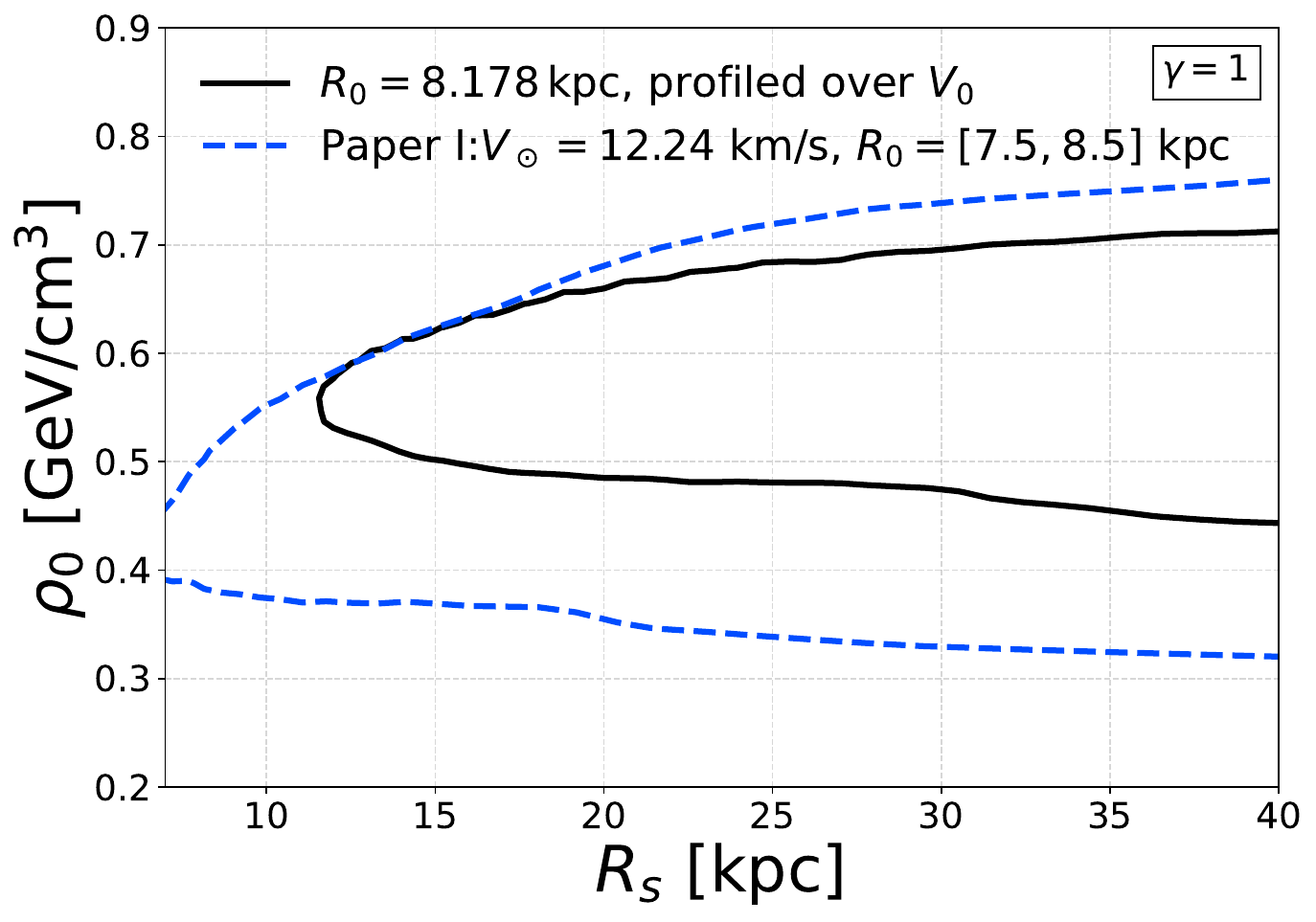}
\includegraphics[scale=0.32]{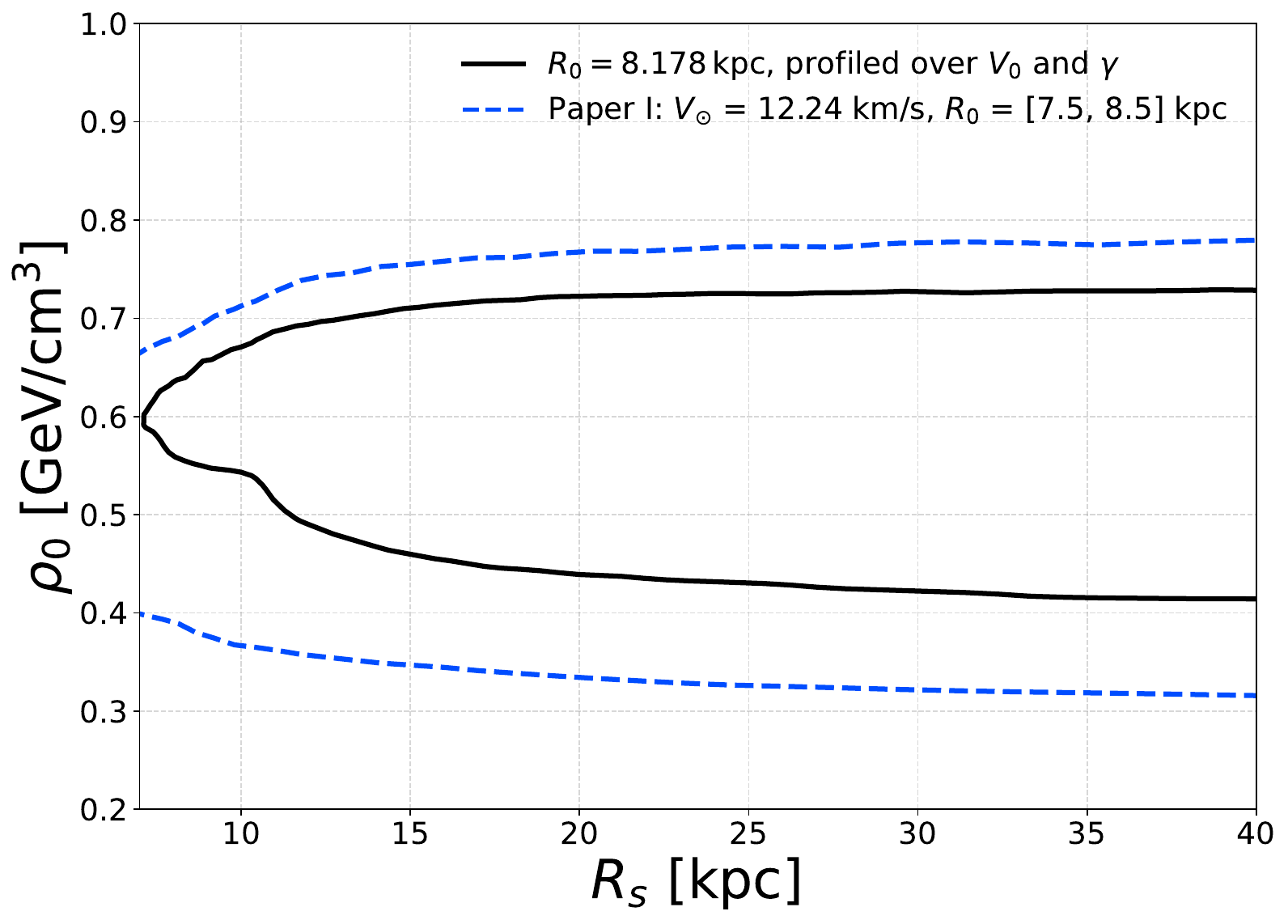}
\caption{Top panel: 2$\sigma$ contours in the $(R_s, \rho_0)$ plane for fixed $\gamma$. The black contour is obtained for fixed $R_0$, and profiled over $V_0$, $\mathcal{M}_i$, $\langle\tau\rangle$ and $\Sigma_*$. The blue dashed line is obtained by profiling over $R_0$, $\mathcal{M}_i$, $\langle\tau\rangle$ and $\Sigma_*$, but for fixed $V_\odot$ (as calculated in Paper I). Bottom panel: same as left panel but further profiled over $V_0$.
}
\label{fig:gNFWcontours_comparison}
\end{figure}

In the top panel of figure~\ref{fig:gNFWcontours_comparison}, we compare the $2\sigma$ contours of the $\chi^2_{\rm RC}$ profiled over $\mathcal{M}_i, \langle\tau\rangle, \Sigma_*$ and $V_0$ (shown in black) -- as obtained in this work --, with the result of paper I, where $V_\odot$ was fixed to 12.24 km/s  and $R_0$ used as independent parameter (see Eq.\ref{eq:sun_angular_velocity}) and varied in the range [7.5,8.5] kpc and the $\chi^2$ profiled over $\mathcal{M}_i, \langle\tau\rangle, \Sigma_*$(blue contour). Both contours are obtained for fixed $\gamma=1$. 
The bottom panel is similar to the top one, but further profiled over $\gamma$. 
The new $R_0$ determination from GRAVITY impacts the constraints on the lower limit of the local DM density, shrinking it by a factor $\sim$30\% in this analysis with respect to those obtained in Paper I. While this improvement is significant, on the other hand is not as dramatic as one might expect given instead the strong improvement in the determination of $R_0$. This is because the uncertainty in $R_0$ is only one of the uncertainties involved in the problem and significant uncertainties still remain, for example in the baryonic morphology, as well as systematics in the determination of the RC.

\subsection{Gaia ranges}
\begin{figure}[h!]
\centering
\includegraphics[scale=0.35]{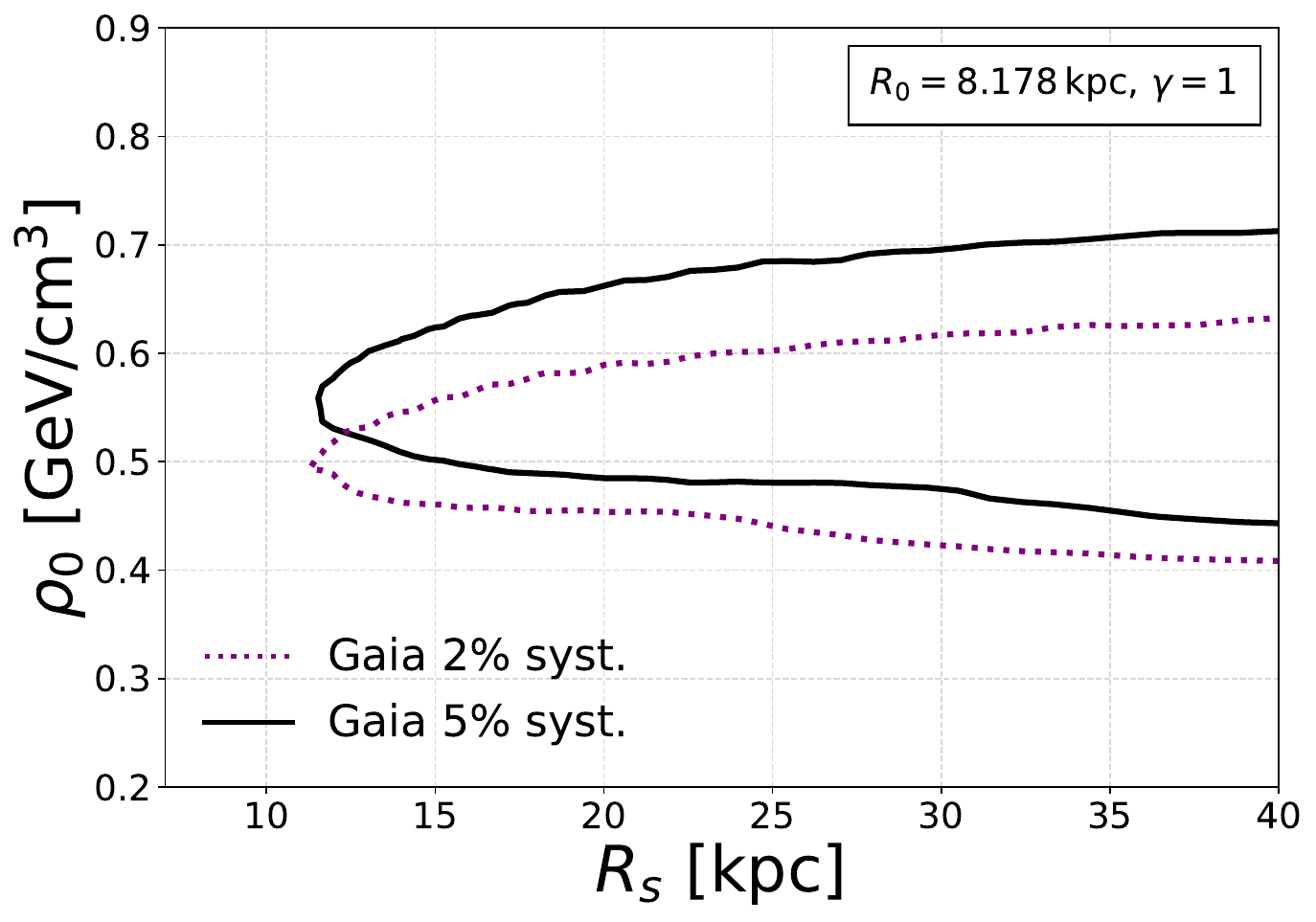}
\includegraphics[scale=0.35]{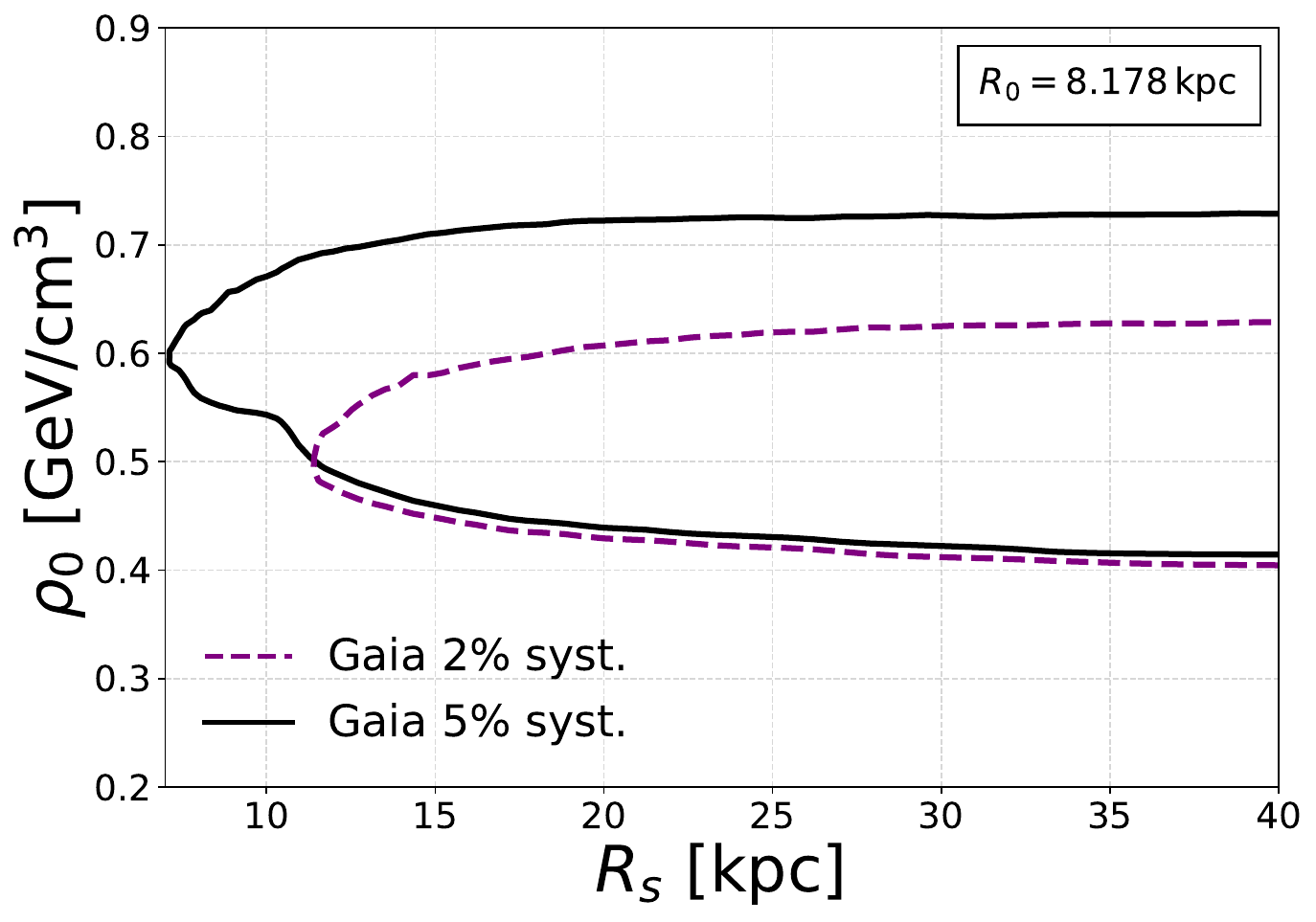}
\caption{Constraints in the $(R_s, \rho_0)$ plane for the two Gaia systematic ranges of $V_0$, as explained in the text. Top panel: 2$\sigma$ contours for fixed $\gamma$, profiled over $V_0$, $\mathcal{M}_i$, $\langle\tau\rangle$ and $\Sigma_*$. Bottom panel:  same as left panel but further profiled over $\gamma$. 
}
\label{fig:gNFWcontours_gaia_comparison}
\end{figure}

Our fiducial range of $V_0$ values, i.e. $[218, 240]\,\rm km/s$, encompass, on the one hand, estimates found in the literature, and, on the other hand, it coincides with the Gaia range estimate assuming a 5\% systematic uncertainty, which is the most pessimistic value considered in \cite{Eilers+19}. If we rather assume a 2\% systematic uncertainty, which is the more optimistic value considered in \cite{Eilers+19}, the Gaia range shrinks to $[224, 234]\,\rm km/s$. 
In figure~\ref{fig:gNFWcontours_gaia_comparison}, we show the constraints obtained in the $(R_s, \rho_0)$ plane for the two Gaia ranges. Although the $V_0$ range is reduced by 50\%, the uncertainty on the local DM density remains virtually unchanged. Similarly to what seen above with $R_0$, this indicates that the uncertainties on the spatial distribution and normalization of baryons  and the large error bars of the RC dominate our determination of the DM distribution in the MW.

\subsection{Comparison with other estimates of $\rho_0$ from the literature}
\label{subsec:rho0_values}

\begin{figure}[h!]
\centering
\includegraphics[scale=0.4]{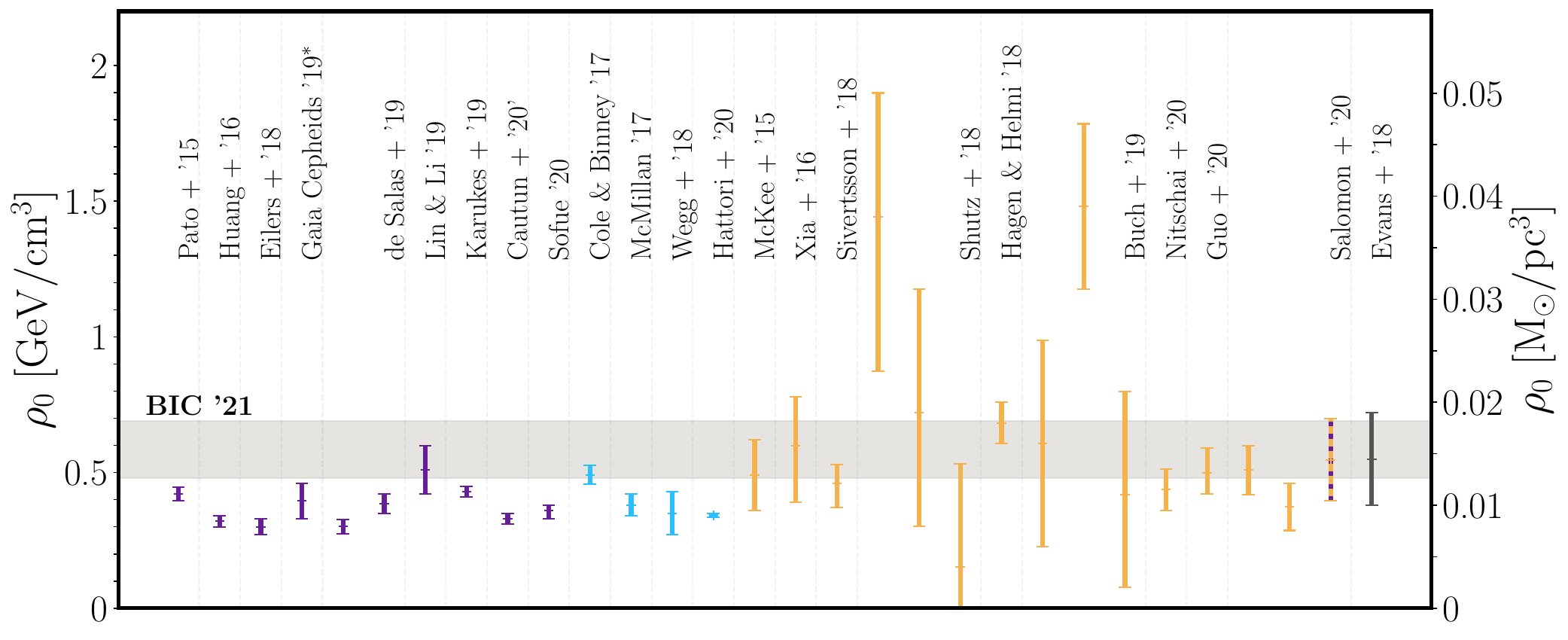}
\caption{Recent determinations of $\rho_0$ as obtained with the RC technique (purple error bars) and other global methods (blue error bars). The yellow error bars correspond to $\rho_0$ values estimated through Jeans modelling of stellar tracers in our Solar neighborhood. The yellow-and-purple error bar is obtained by fitting a global mass model of the Galaxy to local kinematics. Finally, the grey error bar corresponds to the value recommended in the SHM$^{++}$ \cite{2018arXiv181011468E}, the GAIA Cepheids$^*$ datapoint is from the analysis performed here and presented in~\ref{App:cepheidsRC}, and the grey band ``{\tt BIC '21}'' is the main result of this work, using the {\tt galkin} database. Notice that different determinations use different $(R_0, V_0)$ and $(U_\odot, V_\odot, W_\odot)$ values. See text in Section~\ref{subsec:rho0_values} for further details. 
}
\label{fig:rho0_values}
\end{figure}

In figure~\ref{fig:rho0_values} we compare the value of $\rho_0$ obtained in this work (grey band) with other estimates of this parameter as found in the literature. Our inferred local density ranges at the 1$\sigma$ level are as follows:
\begin{equation}
\begin{split}
    \rho_0 &=0.48-\SI{0.67}{GeV/cm^3}\hspace{0.5cm} \textrm{(gNFW)} \\
    \rho_0 &=0.48-\SI{0.67}{GeV/cm^3}\hspace{0.5cm} \textrm{(Einasto)}  \\
    \rho_0 &=0.48-\SI{0.69}{GeV/cm^3}\hspace{0.5cm} \textrm{(Burkert)} .
\end{split}
\end{equation}
Figure~\ref{fig:rho0_values} includes recent values obtained by global fitting of Galactic mass models to the RC \cite{2015ApJ...803L...3P, 2015MNRAS.449..162H, Eilers+19, 2019JCAP...10..037D, 2019MNRAS.487.5679L, 2019JCAP...09..046K, 2020MNRAS.494.4291C, 2020Galax...8...37S} and other techniques, such as fitting of the velocity distribution function, or the application of Jeans equations using global mass models \cite{2017MNRAS.465..798C, 2017MNRAS.465...76M, 2019MNRAS.485.3296W, 2020arXiv201203908H}. We have included the two values of $\rho_0$ as estimated in \cite{2019JCAP...10..037D} adopting two different baryonic mass distributions.
We have also included recent estimates using stellar tracers of the local gravitational force \cite{2015ApJ...814...13M, 2016MNRAS.458.3839X, 2018MNRAS.478.1677S, 2018PhRvL.121h1101S, 2018A&A...615A..99H, 2019JCAP...04..026B, 2020MNRAS.494.6001N, 2020MNRAS.495.4828G, 2020A&A...643A..75S} and the value recommended in the SHM$^{++}$ \cite{2018arXiv181011468E}. 
The three yellow bands for \cite{2018PhRvL.121h1101S} and \cite{2019JCAP...04..026B} correspond to estimates of $\rho_0$ using different populations of stellar tracers. Furthermore, \cite{2020A&A...643A..75S} estimated the local gravitational force using stellar tracers in the Northern and Southern hemisphere, also fitting a global model of the Galaxy to both tracers (yellow-and-purple error bar). For a recent review of techniques and estimates of $\rho_0$, please see \cite{2020arXiv201211477D}.

\clearpage
\section{Conclusions}
\label{ref:conclusions}

We have quantified astrophysical uncertainties on the distribution of Dark Matter in the Milky Way (under the assumption of a gNFW, Burkert and Einasto density profiles) by comparing the observed Rotation Curve with that expected to be caused by the baryonic and DM components of the Galaxy. 
We have made use of state-of-the-art (AD 2020) estimates of the Galactic parameters $(R_0, V_0)$ \cite{Abuter:2018drb, Eilers+19}, updating a previous analysis \cite{2019JCAP...03..033B} also adopting $V_0$ as a new independent variable (instead of $R_0$, as in the previous analysis). 
Our main conclusion is that, despite using the recent precise measurements of $R_0$ and $V_0$ from the Gravity collaboration and Gaia DR2, respectively, 
uncertainties on the determination of the DM distribution stay sizable, and comparable with those estimated with earlier determinations of the Galactic parameters, contrary to general expectations prior to data release. 
This is driven by the fact that the main source of astrophysical uncertainties remains that on the shape and mass of the baryonic component of the Galaxy, and the systematic uncertainties in the observational determination of the Milky Way's rotation curve.

We infer a local density range  $\rho_0$=$0.4-0.7\,{\rm GeV/cm^3}$ at the 2$\sigma$ level, assuming a generalized NFW (gNFW) profile. This range coincides with that obtained under the assumption of an Einasto and a Burkert density profiles, thus indicating that the choice of profile does not affect the determination of local Dark Matter density, within the astrophysical uncertainties.

We provide both the likelihood profile and the Bayesian posterior of the present analysis
-- publicly available at the link in this footnote \footnote{\href{https://github.com/mariabenitocst/UncertaintiesDMinTheMW}{https://github.com/mariabenitocst/UncertaintiesDMinTheMW}}-- so to be adopted in BSM searches to include the most relevant astrophysical uncertainties on the determination of the Dark Matter distribution in the Milky Way.

Adopting state-of-the-art (AD 2021) determinations of Galactic parameters, we find in fact that the uncertainties on quantities relevant for searches of the nature of Dark Matter --propagated from those of astrophysical nature-- are sizable, and should be properly included in all comprehensive analysis.

\clearpage
\appendix
\renewcommand\thefigure{\thesection.\arabic{figure}}  

\section{Burkert profile}
\label{App:burkert_profile}

\begin{figure}[h!]
\centering
\includegraphics[scale=0.35]{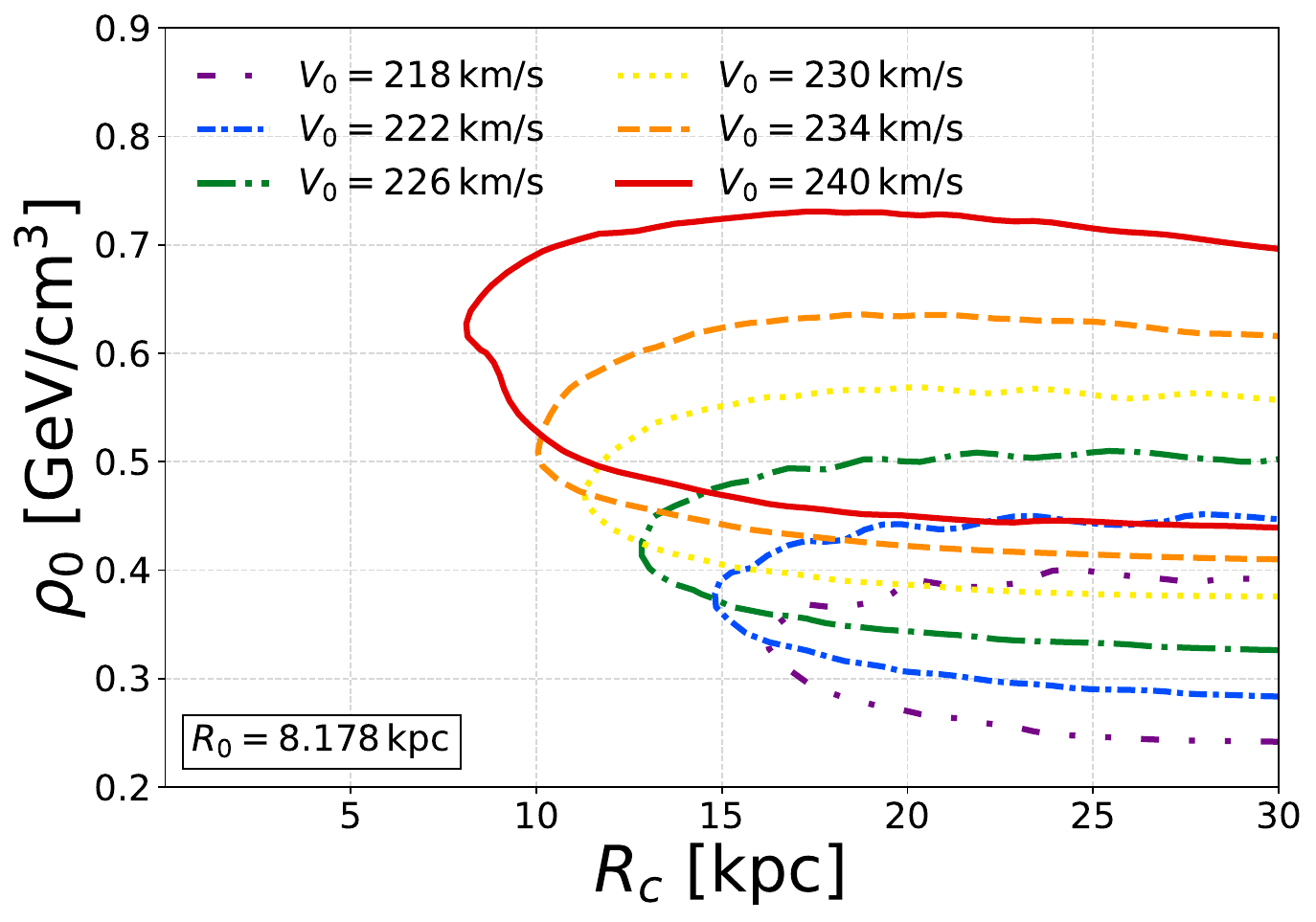}
\includegraphics[scale=0.35]{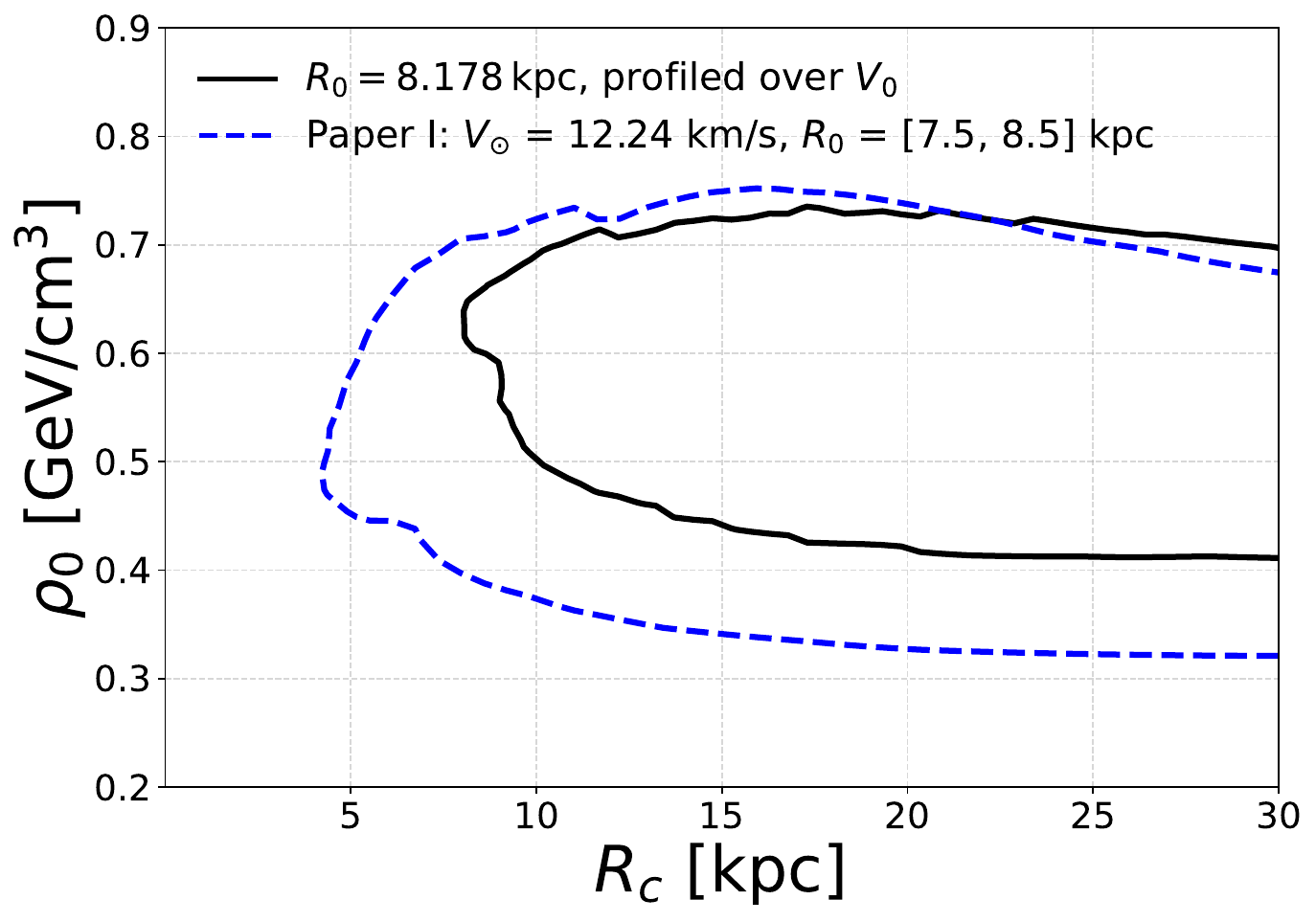}
\caption{2$\sigma$ contours in the ($R_c$, $\rho_0$) plane for a Burkert profile. Top panel: for various values of $V_0$ and profiled over $\mathcal{M}_i$, $\langle\tau\rangle$ and $\Sigma_*$. Bottom panel: further profiled over $V_0$ (black contour), and, in dashed blue, contour obtained in Paper I (profiled over $R_0$, $\mathcal{M}_i$, $\langle\tau\rangle$ and $\Sigma_*$).
}
\label{fig:burkert_contours}
\end{figure}

In this appendix we present the results obtained for a Burkert profile \cite{Burkert_1995}. The Burkert DM density profile has two free parameters: the core radius $R_c$ and the local DM density $\rho_0$. 
In the top panel of figure~\ref{fig:burkert_contours} we present the 2$\sigma$ contours in the $(R_c, \rho_0)$ plane taking into account the latest measurements of the astrophysical quantities $(R_0, V_0)$ \cite{Abuter:2018drb, Eilers+19}. In the bottom panel of the same figure, we compare the 2$\sigma$ contour obtained in this work (i.e. $\chi^2_{\rm RC}$ profiled over $V_0$, $\mathcal{M}_i$, $\langle\tau\rangle$ and $\Sigma_*$) shown in black, with that obtained in Paper I --obtained by profiling over $R_0$, $\mathcal{M}_i$, $\langle\tau\rangle$ and $\Sigma_*$-- which is shown in blue. 
Due to the reduction on uncertainties on astrophysical quantities, the minimum core size is reduced from 5 kpc to roughly 8 kpc.
Furthermore, uncertainties on the local DM density are slightly reduced from $0.33-0.73\,{\rm GeV/cm^3}$ to $0.41-0.73\,{\rm GeV/cm^3}$. As for the gNFW case, uncertainties on our estimate of the DM distribution in the MW are dominated by our ignorance on the actual shape and weight of the baryonic component of the Galaxy. 

\section{Einasto profile}
\label{App:einasto_profile}

\begin{figure}[h!]
\centering
\includegraphics[scale=0.35]{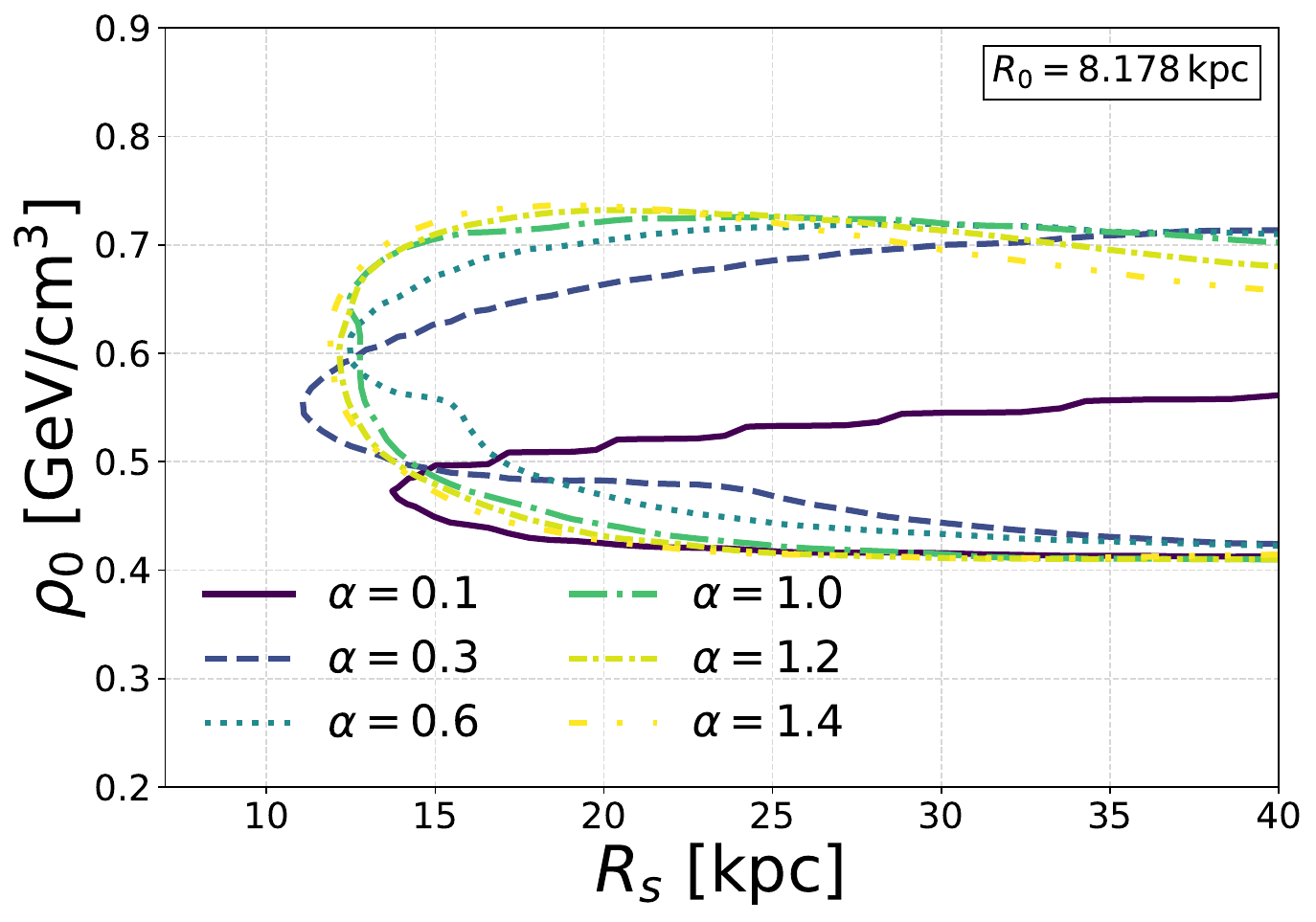}
\includegraphics[scale=0.35]{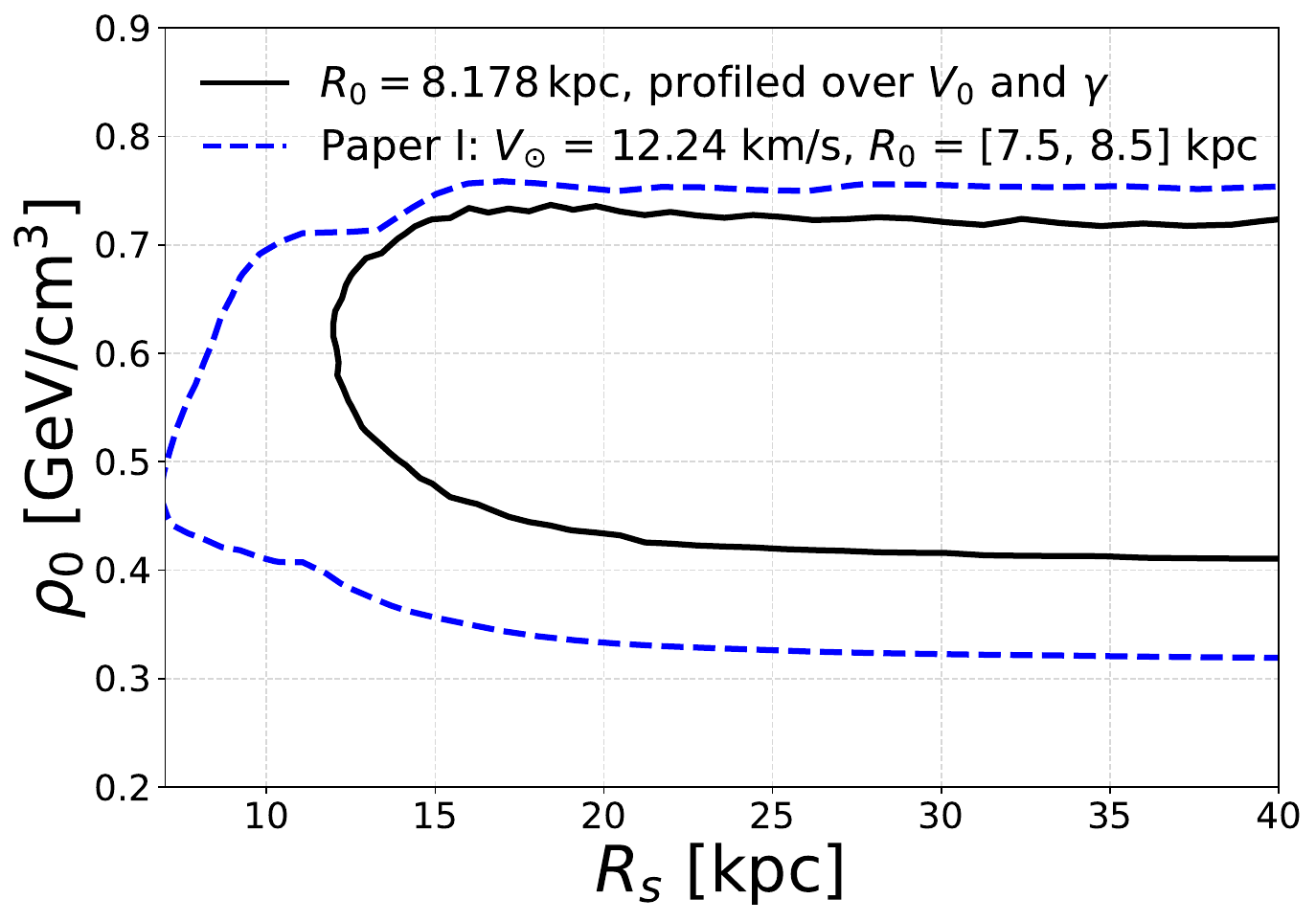}
\caption{Constraints in the $(R_s, \rho_0)$ for an Einasto profile. Left panel: 2$\sigma$ contours for different values of $\alpha$ and fixed $R_0$, profiled over $V_0$, $\mathcal{M}_i$, $\langle\tau\rangle$ and $\Sigma_*$. Right panel: 2$\sigma$ contours for fixed $R_0$, profiled over $\alpha$, $V_0$, $\mathcal{M}_i$, $\langle\tau\rangle$ and $\Sigma_*$ (black contour). The blue contour corresponds to the one obtained in Paper I, i.e. $\chi^2_{RC}$ profiled over $\alpha$, $R_0$, $\mathcal{M}_i$, $\langle\tau\rangle$ and $\Sigma_*$.
}
\label{fig:einasto_contours}
\end{figure}

We also present the results obtained for an Einasto DM density profile \cite{1965TrAlm...5...87E}, which is defined in terms of the shape parameter (or inner slope of the logarithmic density profile) $\alpha$, the scale radius $R_s$ and the local DM density $\rho_0$.
The left panel of figure~\ref{fig:einasto_contours} shows the $2\sigma$ contours obtained in the ($R_s$, $\rho_0$) plane,  for different values of the parameter $\alpha$ and profiled over $V_0$, $\mathcal{M}_i$, $\langle\tau\rangle$ and $\Sigma_0$, while taking into account the recent estimations of the Sun's galactocentric distance and its circular velocity \cite{Abuter:2018drb, Eilers+19}. The right panel of figure~\ref{fig:einasto_contours} compares the constraints obtained in light of new astrophysical data (black contour) with the results obtained in Paper I (blue contour). In light of new estimates of the Sun's distance to the GC and its circular velocity, the allowed 2$\sigma$ range for the local DM density is $0.41-0.73\,{\rm GeV/cm^3}$.

\section{Bayesian framework}
\label{App:bayesian}

In this section we present the results of a fully Bayesian analysis. By comparing the results obtained with the frequentist and Bayesian frameworks, we are able to bracket uncertainties due to the use of the statistical methodology.
For a given baryonic morphology, our model has six free parameters: the three parameters $(R_s, \rho_0, \gamma)$ of the gNFW density profile, the two parameters that control the normalization of the baryonic mass, namely $\langle\tau\rangle$ and $\Sigma_*$, and the Sun's circular velocity $V_0$.
We perform a Monte Carlo scan of the parameter space by means of the nested sampling code {\tt PyMultiNest} \citep{2009MNRAS.398.1601F, 2014A&A...564A.125B}, using flat priors on the parameters.
We account for the uncertainty in the choice of baryonic morphology by repeating the scan for each different morphology and then performing a Bayesian model averaging (e.g. \cite{2008ConPh..49...71T}). 
In particular, we follow the prescription described in section  2.4.2 of \cite{2019arXiv191204296K}, with the only difference that, in the analysis presented here, $V_0$ is a free parameter. 
In short, this means that 30 different six-dimensional posterior distributions are calculated, one for each baryonic morphology, and then they are averaged to get the final one.
The six-dimensional model-averaged posterior can be found at \href{https://github.com/mariabenitocst/UncertaintiesDMinTheMW}{https://github.com/mariabenitocst/UncertaintiesDMinTheMW}.

\begin{figure}[h!]
\centering
\includegraphics[scale=0.45]{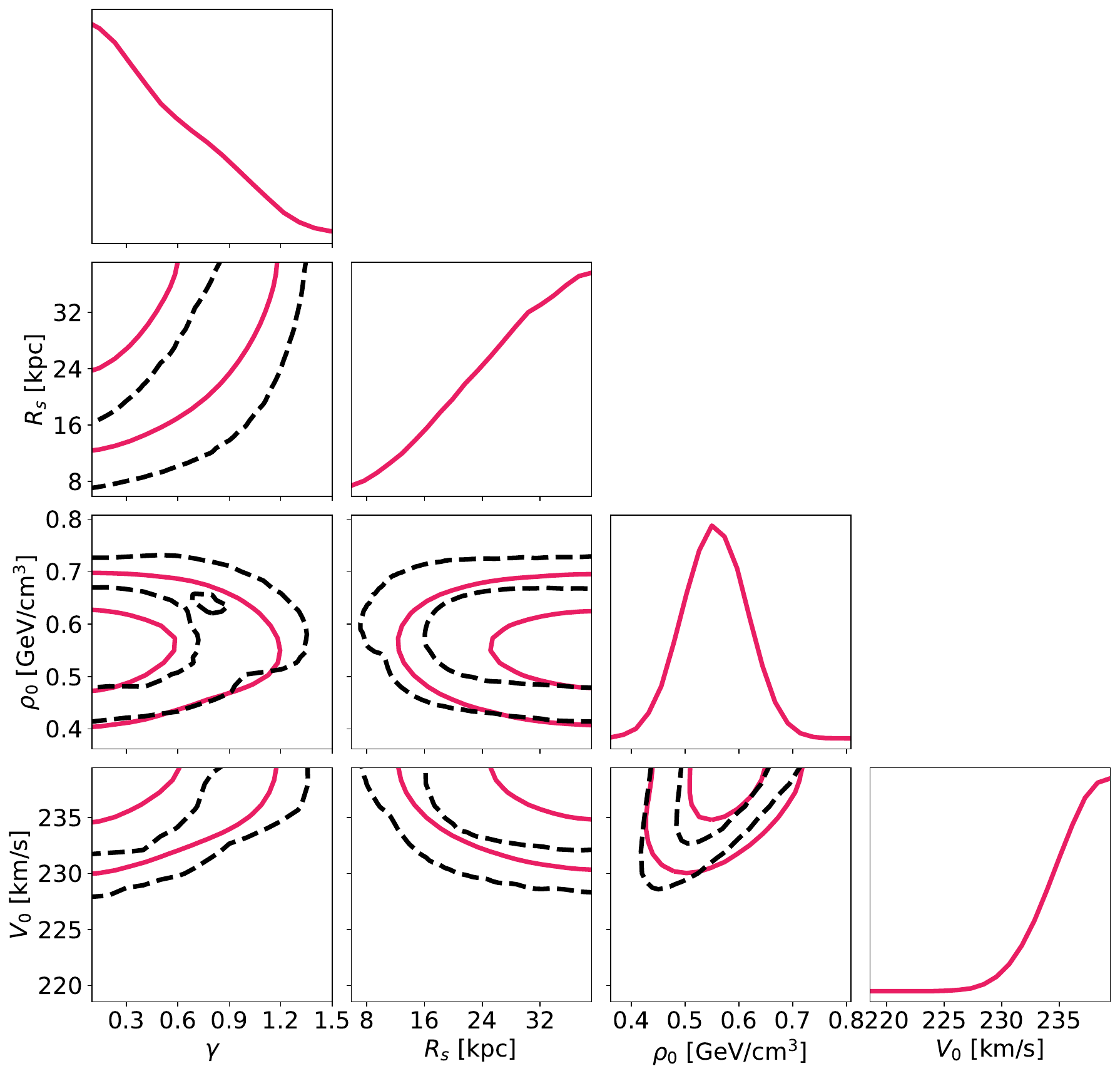}
\caption{One and two-dimensional marginalized Bayesian posterior distributions for the  baryonic model-averaged case (magenta). The Bayesian contours delimit regions of 68\% and 95\% probability.
The frequentist contours delimiting the 68\% and 95\% confidence regions are also shown in black.
The full posterior is 6-d, but for better clearness, we only show the triangle plot in the four most relevant parameters, i.e., $R_s, \rho_0, \gamma$ and $V_0$. 
}
\label{fig:posterior}
\end{figure}

Figure \ref{fig:posterior} shows the one and two-dimensional marginalized posterior distributions for the model-averaged. The Bayesian contours (shown in magenta) delimiting regions of 68\% and 95\% probability are compared with the 1-2 $\sigma$ frequentist contours, which are shown in black.
The Bayesian model-averaged contours are less conservative than the frequentist counterparts and thus, as observed in the figure, the former contours are smaller. 

\section{Cepheids Gaia Rotation Curve}
\label{App:cepheidsRC}

In \cite{2019ApJ...870L..10M}, the authors obtain the RC between 4 and 20 kpc from the Galactic center using classical Cepheids --with proper motions and radial velocities measured by Gaia DR2-- as tracers. 
We wish to investigate the constraints in the Galactic distribution of DM set by this new data, adopting the RC in the data format presented in the aforementioned analysis as it permits the same binning procedure we used for the {\tt galkin} compilation. 
For this check, we adopt the Cepheids Gaia RC assuming $R_0=\SI{8.09}{kpc}$ and $V_0=\SI{233.6}{km/s}$, which are the values estimated in \cite{2019ApJ...870L..10M}. We first bin the RC and then, we perform a scan in the 6-dimensional parameter space $(\gamma, r_s, \rho_0, \mathcal{M}_i, \langle\tau\rangle, \Sigma_*)$ closely following the procedure described in section~\ref{sec:meth}. 

\begin{figure}[h!]
\centering
\includegraphics[scale=0.45]{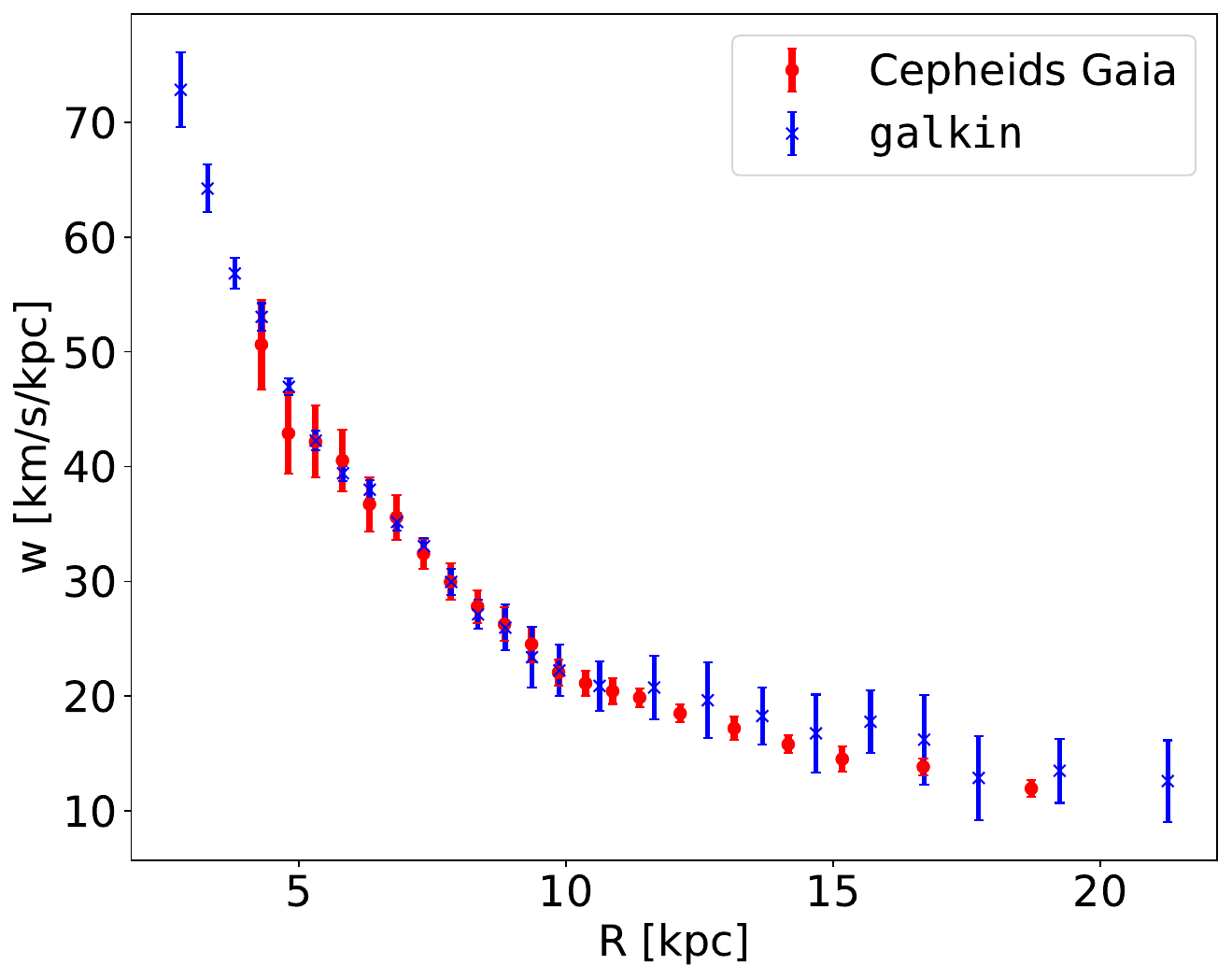}
\caption{Binned angular rotation curve for the Cepheids Gaia and {\tt galkin} data sets with fixed $R_0=\SI{8.09}{kpc}$ and $V_0=\SI{233.6}{km/s}$.
}
\label{fig:cepheidsRC}
\end{figure}

\begin{figure}[h!]
\centering
\includegraphics[scale=0.3]{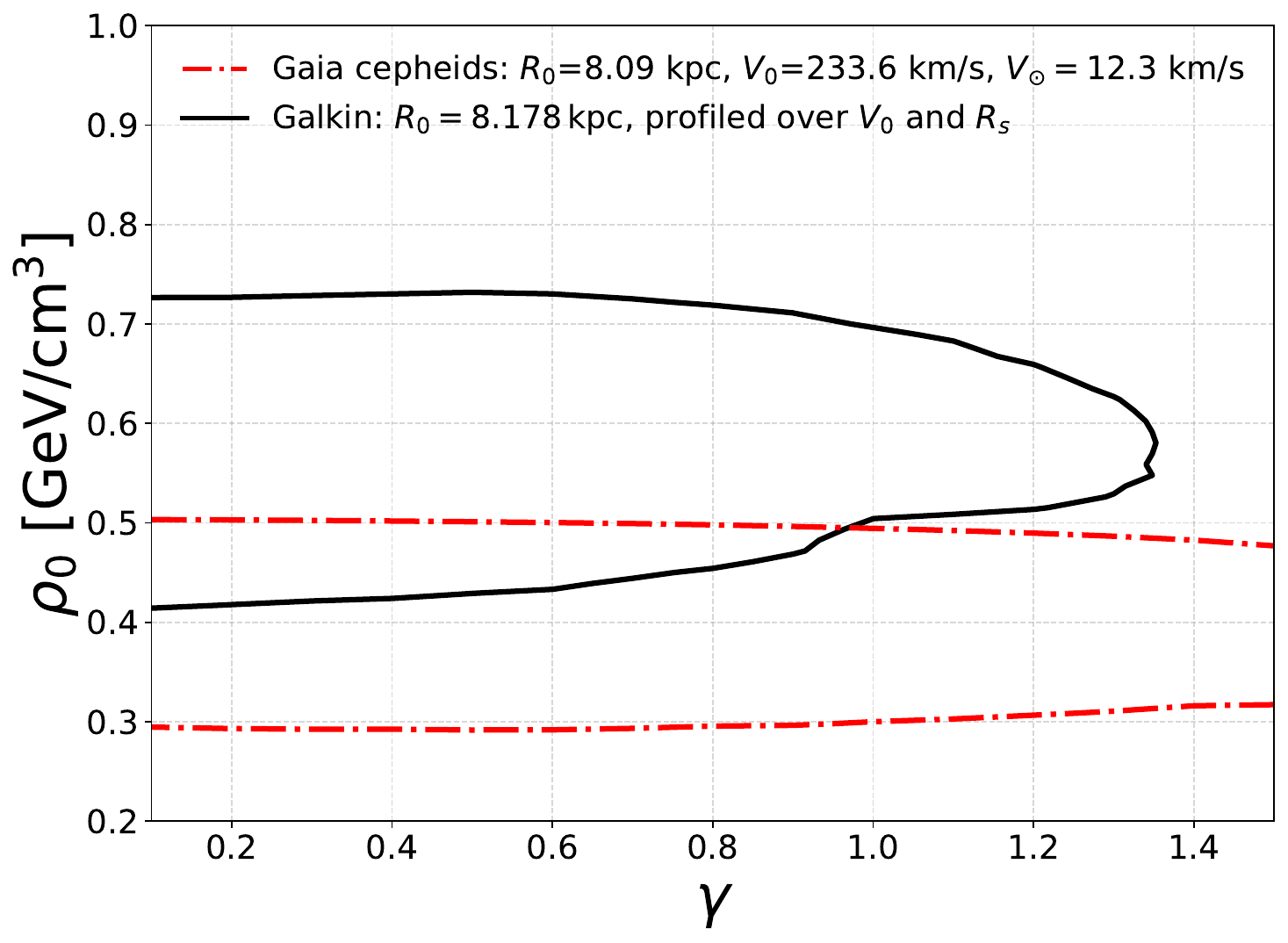}
\includegraphics[scale=0.3]{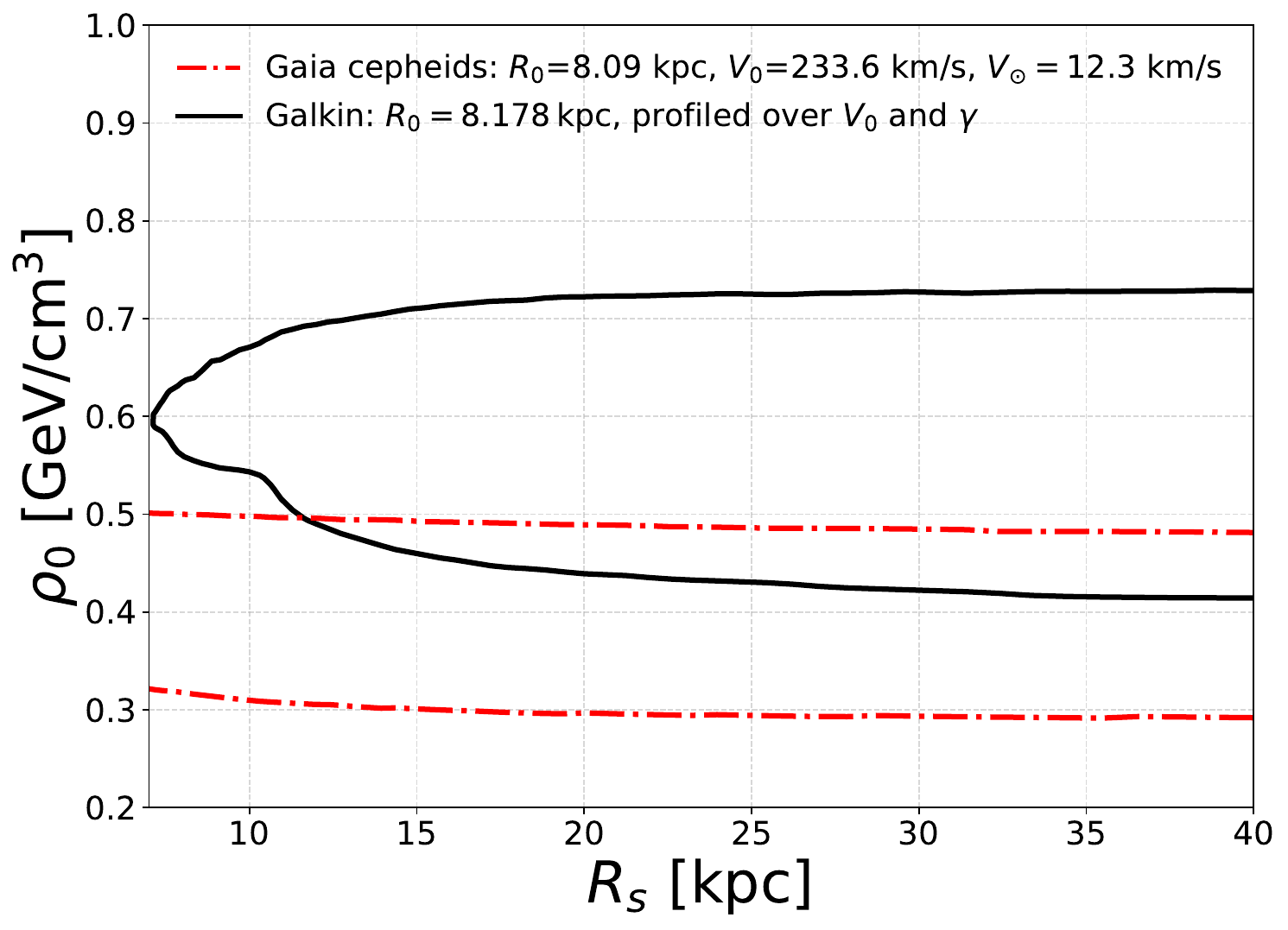}
\caption{2$\sigma$ contours in the $(\gamma, \rho_0)$ (top panel) and $(R_s, \rho_0)$ (bottom panel) planes for the Cepheids Gaia RC (dash-dotted red) and the {\tt galkin} data set as obtained in this work (solid black). Top panel: contours are profiled over baryonic morphology, normalization and $R_s$. The contours obtained for the {\tt galkin} dataset are further profiled over $V_0$. 
}
\label{fig:cepheids_contours}
\end{figure}

Figure~\ref{fig:cepheidsRC} compares the binned RC as obtained from the {\tt galkin} data set and Cepheids Gaia. Figure~\ref{fig:cepheids_contours} compares the 2$\sigma$ contours obtained for the Cepheids RC with those obtained for the {\tt galkin} data set. It can be seen that the Cepheids data seem to prefer $\rho_0$ values slightly smaller than those preferred by the {\tt galkin} compilation, while still being in full agreement with each other. It is to be noticed that the Gaia contours are  only for a fixed value of $V_0$, thus the smaller region obtained should not generate surprise.

\bibliographystyle{elsarticle-num} 
\bibliography{main}

\begin{thebibliography}{10}
\expandafter\ifx\csname url\endcsname\relax
  \def\url#1{\texttt{#1}}\fi
\expandafter\ifx\csname urlprefix\endcsname\relax\def\urlprefix{URL }\fi
\expandafter\ifx\csname href\endcsname\relax
  \def\href#1#2{#2} \def\path#1{#1}\fi

\bibitem{2015NatPh..11..245I}
F.~{Iocco}, M.~{Pato}, G.~{Bertone}, {Evidence for dark matter in the inner
  Milky Way}, Nature Physics 11~(3) (2015) 245--248.
\newblock \href {http://arxiv.org/abs/1502.03821} {\path{arXiv:1502.03821}},
  \href {https://doi.org/10.1038/nphys3237} {\path{doi:10.1038/nphys3237}}.

\bibitem{2015ApJ...803L...3P}
M.~{Pato}, F.~{Iocco}, {The Dark Matter Profile of the Milky Way: A
  Non-parametric Reconstruction}, \apjl 803~(1) (2015) L3.
\newblock \href {http://arxiv.org/abs/1504.03317} {\path{arXiv:1504.03317}},
  \href {https://doi.org/10.1088/2041-8205/803/1/L3}
  {\path{doi:10.1088/2041-8205/803/1/L3}}.

\bibitem{2013ApJ...779..115B}
J.~{Bovy}, H.-W. {Rix}, {A Direct Dynamical Measurement of the Milky Way's Disk
  Surface Density Profile, Disk Scale Length, and Dark Matter Profile at 4 kpc
  \&lt;\raisebox{-0.5ex}\textasciitilde R \&lt;\raisebox{-0.5ex}\textasciitilde
  9 kpc}, \apj 779~(2) (2013) 115.
\newblock \href {http://arxiv.org/abs/1309.0809} {\path{arXiv:1309.0809}},
  \href {https://doi.org/10.1088/0004-637X/779/2/115}
  {\path{doi:10.1088/0004-637X/779/2/115}}.

\bibitem{2015JCAP...12..001P}
M.~{Pato}, F.~{Iocco}, G.~{Bertone}, {Dynamical constraints on the dark matter
  distribution in the Milky Way}, \jcap 2015~(12) (2015) 001.
\newblock \href {http://arxiv.org/abs/1504.06324} {\path{arXiv:1504.06324}},
  \href {https://doi.org/10.1088/1475-7516/2015/12/001}
  {\path{doi:10.1088/1475-7516/2015/12/001}}.

\bibitem{2017PDU....15...90I}
F.~{Iocco}, M.~{Benito}, {An estimate of the DM profile in the Galactic bulge
  region}, Physics of the Dark Universe 15 (2017) 90--95.
\newblock \href {http://arxiv.org/abs/1611.09861} {\path{arXiv:1611.09861}},
  \href {https://doi.org/10.1016/j.dark.2016.12.004}
  {\path{doi:10.1016/j.dark.2016.12.004}}.

\bibitem{2017JCAP...02..007B}
M.~{Benito}, N.~{Bernal}, N.~{Bozorgnia}, F.~{Calore}, F.~{Iocco}, {Particle
  Dark Matter constraints: the effect of Galactic uncertainties}, \jcap
  2017~(2) (2017) 007.
\newblock \href {http://arxiv.org/abs/1612.02010} {\path{arXiv:1612.02010}},
  \href {https://doi.org/10.1088/1475-7516/2017/02/007}
  {\path{doi:10.1088/1475-7516/2017/02/007}}.

\bibitem{2019arXiv191204296K}
E.~V. {Karukes}, M.~{Benito}, F.~{Iocco}, R.~{Trotta}, A.~{Geringer-Sameth}, {A
  robust estimate of the Milky Way mass from rotation curve data}, \jcap
  2020~(5) (2020) 033.
\newblock \href {http://arxiv.org/abs/1912.04296} {\path{arXiv:1912.04296}},
  \href {https://doi.org/10.1088/1475-7516/2020/05/033}
  {\path{doi:10.1088/1475-7516/2020/05/033}}.

\bibitem{2019JCAP...09..046K}
E.~V. {Karukes}, M.~{Benito}, F.~{Iocco}, R.~{Trotta}, A.~{Geringer-Sameth},
  {Bayesian reconstruction of the Milky Way dark matter distribution}, \jcap
  2019~(9) (2019) 046.
\newblock \href {http://arxiv.org/abs/1901.02463} {\path{arXiv:1901.02463}},
  \href {https://doi.org/10.1088/1475-7516/2019/09/046}
  {\path{doi:10.1088/1475-7516/2019/09/046}}.

\bibitem{Nesti:2013uwa}
F.~Nesti, P.~Salucci, {The Dark Matter halo of the Milky Way, AD 2013}, JCAP
  1307 (2013) 016.
\newblock \href {http://arxiv.org/abs/1304.5127} {\path{arXiv:1304.5127}},
  \href {https://doi.org/10.1088/1475-7516/2013/07/016}
  {\path{doi:10.1088/1475-7516/2013/07/016}}.

\bibitem{Silverwood:2015hxa}
H.~Silverwood, S.~Sivertsson, P.~Steger, J.~I. Read, G.~Bertone, {A
  non-parametric method for measuring the local dark matter density}, Mon. Not.
  Roy. Astron. Soc. 459~(4) (2016) 4191--4208.
\newblock \href {http://arxiv.org/abs/1507.08581} {\path{arXiv:1507.08581}},
  \href {https://doi.org/10.1093/mnras/stw917}
  {\path{doi:10.1093/mnras/stw917}}.

\bibitem{Catena:2009mf}
R.~Catena, P.~Ullio, {A novel determination of the local dark matter density},
  JCAP 1008 (2010) 004.
\newblock \href {http://arxiv.org/abs/0907.0018} {\path{arXiv:0907.0018}},
  \href {https://doi.org/10.1088/1475-7516/2010/08/004}
  {\path{doi:10.1088/1475-7516/2010/08/004}}.

\bibitem{2018MNRAS.478.1677S}
S.~{Sivertsson}, H.~{Silverwood}, J.~I. {Read}, G.~{Bertone}, P.~{Steger}, {The
  local dark matter density from SDSS-SEGUE G-dwarfs}, \mnras 478~(2) (2018)
  1677--1693.
\newblock \href {http://arxiv.org/abs/1708.07836} {\path{arXiv:1708.07836}},
  \href {https://doi.org/10.1093/mnras/sty977}
  {\path{doi:10.1093/mnras/sty977}}.

\bibitem{2019JCAP...10..037D}
P.~F. {de Salas}, K.~{Malhan}, K.~{Freese}, K.~{Hattori}, M.~{Valluri}, {On the
  estimation of the local dark matter density using the rotation curve of the
  Milky Way}, \jcap 2019~(10) (2019) 037.
\newblock \href {http://arxiv.org/abs/1906.06133} {\path{arXiv:1906.06133}},
  \href {https://doi.org/10.1088/1475-7516/2019/10/037}
  {\path{doi:10.1088/1475-7516/2019/10/037}}.

\bibitem{2020MNRAS.tmp.1873W}
A.~{Widmark}, K.~{Malhan}, P.~F. {de Salas}, S.~{Sivertsson}, {Measuring the
  Matter Density of the Galactic Disk Using Stellar Streams}, \mnras (Jun.
  2020).
\newblock \href {http://arxiv.org/abs/2003.04318} {\path{arXiv:2003.04318}},
  \href {https://doi.org/10.1093/mnras/staa1741}
  {\path{doi:10.1093/mnras/staa1741}}.

\bibitem{2020arXiv201211477D}
P.~F. {de Salas}, A.~{Widmark}, {Dark matter local density determination:
  recent observations and future prospects}, arXiv e-prints (2020)
  arXiv:2012.11477\href {http://arxiv.org/abs/2012.11477}
  {\path{arXiv:2012.11477}}.

\bibitem{2019JCAP...03..033B}
M.~{Benito}, A.~{Cuoco}, F.~{Iocco}, {Handling the uncertainties in the
  Galactic Dark Matter distribution for particle Dark Matter searches}, \jcap
  2019~(3) (2019) 033.
\newblock \href {http://arxiv.org/abs/1901.02460} {\path{arXiv:1901.02460}},
  \href {https://doi.org/10.1088/1475-7516/2019/03/033}
  {\path{doi:10.1088/1475-7516/2019/03/033}}.

\bibitem{2019JCAP...04..040F}
Y.~{Farzan}, M.~{Rajaee}, {Dark matter decaying into millicharged particles as
  a solution to AMS-02 positron excess}, \jcap 2019~(4) (2019) 040.
\newblock \href {http://arxiv.org/abs/1901.11273} {\path{arXiv:1901.11273}},
  \href {https://doi.org/10.1088/1475-7516/2019/04/040}
  {\path{doi:10.1088/1475-7516/2019/04/040}}.

\bibitem{2019PhRvD.100d3020A}
J.~F. {Acevedo}, J.~{Bramante}, {Supernovae sparked by dark matter in white
  dwarfs}, \prd 100~(4) (2019) 043020.
\newblock \href {http://arxiv.org/abs/1904.11993} {\path{arXiv:1904.11993}},
  \href {https://doi.org/10.1103/PhysRevD.100.043020}
  {\path{doi:10.1103/PhysRevD.100.043020}}.

\bibitem{2019PhRvD.100j3014L}
S.-J. {Lin}, X.-J. {Bi}, P.-F. {Yin}, {Investigating the dark matter signal in
  the cosmic ray antiproton flux with the machine learning method}, \prd
  100~(10) (2019) 103014.
\newblock \href {http://arxiv.org/abs/1903.09545} {\path{arXiv:1903.09545}},
  \href {https://doi.org/10.1103/PhysRevD.100.103014}
  {\path{doi:10.1103/PhysRevD.100.103014}}.

\bibitem{2019arXiv191209486A}
C.~A. {Arg{\"u}elles}, A.~{Diaz}, A.~{Kheirandish}, et~al., {Dark Matter
  Annihilation to Neutrinos: An Updated, Consistent \&amp; Compelling
  Compendium of Constraints}, arXiv e-prints (2019) arXiv:1912.09486\href
  {http://arxiv.org/abs/1912.09486} {\path{arXiv:1912.09486}}.

\bibitem{2020PhRvD.101d3526S}
J.~S. {Sidhu}, {Charge constraints of macroscopic dark matter}, \prd 101~(4)
  (2020) 043526.
\newblock \href {http://arxiv.org/abs/1912.04732} {\path{arXiv:1912.04732}},
  \href {https://doi.org/10.1103/PhysRevD.101.043526}
  {\path{doi:10.1103/PhysRevD.101.043526}}.

\bibitem{2020arXiv200306614A}
{ANTARES Collaboration}, A.~{Albert}, M.~{Andr{\'e}}, M.~{Anghinolfi}, et~al.,
  {Combined search for neutrinos from dark matter self-annihilation in the
  Galactic Centre with ANTARES and IceCube}, arXiv e-prints (2020)
  arXiv:2003.06614\href {http://arxiv.org/abs/2003.06614}
  {\path{arXiv:2003.06614}}.

\bibitem{2020arXiv200508824B}
G.~{Baym}, D.~H. {Beck}, J.~P. {Filippini}, C.~J. {Pethick}, J.~{Shelton},
  {Searching for low mass dark matter via phonon creation in superfluid 4He},
  arXiv e-prints (2020) arXiv:2005.08824\href {http://arxiv.org/abs/2005.08824}
  {\path{arXiv:2005.08824}}.

\bibitem{2020arXiv200606836G}
G.~B. {Gelmini}, A.~J. {Millar}, V.~{Takhistov}, E.~{Vitagliano}, {Probing Dark
  Photons with Plasma Haloscopes}, arXiv e-prints (2020) arXiv:2006.06836\href
  {http://arxiv.org/abs/2006.06836} {\path{arXiv:2006.06836}}.

\bibitem{2019PhRvL.122q1801B}
T.~{Bringmann}, M.~{Pospelov}, {Novel Direct Detection Constraints on Light
  Dark Matter}, \prl 122~(17) (2019) 171801.
\newblock \href {http://arxiv.org/abs/1810.10543} {\path{arXiv:1810.10543}},
  \href {https://doi.org/10.1103/PhysRevLett.122.171801}
  {\path{doi:10.1103/PhysRevLett.122.171801}}.

\bibitem{2017SoftX...6...54P}
M.~{Pato}, F.~{Iocco}, {galkin: A new compilation of Milky Way rotation curve
  data}, SoftwareX 6 (2017) 54--62.
\newblock \href {http://arxiv.org/abs/1703.00020} {\path{arXiv:1703.00020}},
  \href {https://doi.org/10.1016/j.softx.2016.12.006}
  {\path{doi:10.1016/j.softx.2016.12.006}}.

\bibitem{DrimmelPoggio18}
R.~{Drimmel}, E.~{Poggio}, {On the Solar Velocity}, Research Notes of the
  American Astronomical Society 2~(4) (2018) 210.
\newblock \href {https://doi.org/10.3847/2515-5172/aaef8b}
  {\path{doi:10.3847/2515-5172/aaef8b}}.

\bibitem{1996MNRAS.278..488Z}
H.~{Zhao}, {Analytical models for galactic nuclei}, \mnras 278~(2) (1996)
  488--496.
\newblock \href {http://arxiv.org/abs/astro-ph/9509122}
  {\path{arXiv:astro-ph/9509122}}, \href
  {https://doi.org/10.1093/mnras/278.2.488}
  {\path{doi:10.1093/mnras/278.2.488}}.

\bibitem{2001ApJ...555..504W}
J.~S.~B. {Wyithe}, E.~L. {Turner}, D.~N. {Spergel}, {Gravitational Lens
  Statistics for Generalized NFW Profiles: Parameter Degeneracy and
  Implications for Self-Interacting Cold Dark Matter}, \apj 555~(1) (2001)
  504--523.
\newblock \href {http://arxiv.org/abs/astro-ph/0007354}
  {\path{arXiv:astro-ph/0007354}}, \href {https://doi.org/10.1086/321437}
  {\path{doi:10.1086/321437}}.

\bibitem{Burkert_1995}
A.~Burkert, \href{http://dx.doi.org/10.1086/309560}{The structure of dark
  matter halos in dwarf galaxies}, The Astrophysical Journal 447~(1) (Jul
  1995).
\newblock \href {https://doi.org/10.1086/309560} {\path{doi:10.1086/309560}}.
\newline\urlprefix\url{http://dx.doi.org/10.1086/309560}

\bibitem{1965TrAlm...5...87E}
J.~{Einasto}, {On the Construction of a Composite Model for the Galaxy and on
  the Determination of the System of Galactic Parameters}, Trudy
  Astrofizicheskogo Instituta Alma-Ata 5 (1965) 87--100.

\bibitem{2005ApJ...631..879P}
P.~{Popowski}, K.~{Griest}, C.~L. {Thomas}, et~al., {Microlensing Optical Depth
  toward the Galactic Bulge Using Clump Giants from the MACHO Survey}, \apj
  631~(2) (2005) 879--905.
\newblock \href {http://arxiv.org/abs/astro-ph/0410319}
  {\path{arXiv:astro-ph/0410319}}, \href {https://doi.org/10.1086/432246}
  {\path{doi:10.1086/432246}}.

\bibitem{GRAVITY2019}
{Gravity Collaboration}, R.~{Abuter}, A.~{Amorim}, M.~{Baub{\"o}ck}, et~al., {A
  geometric distance measurement to the Galactic center black hole with 0.3\%
  uncertainty}, \aap 625 (2019) L10.
\newblock \href {http://arxiv.org/abs/1904.05721} {\path{arXiv:1904.05721}},
  \href {https://doi.org/10.1051/0004-6361/201935656}
  {\path{doi:10.1051/0004-6361/201935656}}.

\bibitem{2016A&A...595A...1G}
{Gaia Collaboration}, T.~{Prusti}, J.~H.~J. {de Bruijne}, A.~G.~A. {Brown},
  et~al., {The Gaia mission}, \aap 595 (2016) A1.
\newblock \href {http://arxiv.org/abs/1609.04153} {\path{arXiv:1609.04153}},
  \href {https://doi.org/10.1051/0004-6361/201629272}
  {\path{doi:10.1051/0004-6361/201629272}}.

\bibitem{2010AJ....140.1868W}
E.~L. {Wright}, P.~R.~M. {Eisenhardt}, A.~K. {Mainzer}, et~al., {The Wide-field
  Infrared Survey Explorer (WISE): Mission Description and Initial On-orbit
  Performance}, \aj 140~(6) (2010) 1868--1881.
\newblock \href {http://arxiv.org/abs/1008.0031} {\path{arXiv:1008.0031}},
  \href {https://doi.org/10.1088/0004-6256/140/6/1868}
  {\path{doi:10.1088/0004-6256/140/6/1868}}.

\bibitem{2006AJ....131.1163S}
M.~F. {Skrutskie}, R.~M. {Cutri}, R.~{Stiening}, et~al., {The Two Micron All
  Sky Survey (2MASS)}, \aj 131~(2) (2006) 1163--1183.
\newblock \href {https://doi.org/10.1086/498708} {\path{doi:10.1086/498708}}.

\bibitem{2017AJ....154...94M}
S.~R. {Majewski}, R.~P. {Schiavon}, P.~M. {Frinchaboy}, et~al., {The Apache
  Point Observatory Galactic Evolution Experiment (APOGEE)}, \aj 154~(3) (2017)
  94.
\newblock \href {http://arxiv.org/abs/1509.05420} {\path{arXiv:1509.05420}},
  \href {https://doi.org/10.3847/1538-3881/aa784d}
  {\path{doi:10.3847/1538-3881/aa784d}}.

\bibitem{Eilers+19}
A.-C. {Eilers}, D.~W. {Hogg}, H.-W. {Rix}, M.~K. {Ness}, {The Circular Velocity
  Curve of the Milky Way from 5 to 25 kpc}, \apj 871~(1) (2019) 120.
\newblock \href {http://arxiv.org/abs/1810.09466} {\path{arXiv:1810.09466}},
  \href {https://doi.org/10.3847/1538-4357/aaf648}
  {\path{doi:10.3847/1538-4357/aaf648}}.

\bibitem{GRAVITY2020}
{GRAVITY Collaboration}, R.~{Abuter}, A.~{Amorim}, M.~{Bauboeck}, et~al.,
  {Detection of the Schwarzschild precession in the orbit of the star S2 near
  the Galactic centre massive black hole}, arXiv e-prints (2020)
  arXiv:2004.07187\href {http://arxiv.org/abs/2004.07187}
  {\path{arXiv:2004.07187}}.

\bibitem{2012ApJ...759..131B}
J.~{Bovy}, C.~{Allende Prieto}, T.~C. {Beers}, et~al., {The Milky Way's
  Circular-velocity Curve between 4 and 14 kpc from APOGEE data}, \apj 759~(2)
  (2012) 131.
\newblock \href {http://arxiv.org/abs/1209.0759} {\path{arXiv:1209.0759}},
  \href {https://doi.org/10.1088/0004-637X/759/2/131}
  {\path{doi:10.1088/0004-637X/759/2/131}}.

\bibitem{2014ApJ...783..130R}
M.~J. {Reid}, K.~M. {Menten}, A.~{Brunthaler}, et~al., {Trigonometric
  Parallaxes of High Mass Star Forming Regions: The Structure and Kinematics of
  the Milky Way}, \apj 783~(2) (2014) 130.
\newblock \href {http://arxiv.org/abs/1401.5377} {\path{arXiv:1401.5377}},
  \href {https://doi.org/10.1088/0004-637X/783/2/130}
  {\path{doi:10.1088/0004-637X/783/2/130}}.

\bibitem{BlandHawthornGerhard16}
J.~{Bland-Hawthorn}, O.~{Gerhard}, {The Galaxy in Context: Structural,
  Kinematic, and Integrated Properties}, \araa 54 (2016) 529--596.
\newblock \href {http://arxiv.org/abs/1602.07702} {\path{arXiv:1602.07702}},
  \href {https://doi.org/10.1146/annurev-astro-081915-023441}
  {\path{doi:10.1146/annurev-astro-081915-023441}}.

\bibitem{Schonrich2012}
R.~{Sch{\"o}nrich}, {Galactic rotation and solar motion from stellar
  kinematics}, \mnras 427~(1) (2012) 274--287.
\newblock \href {http://arxiv.org/abs/1207.3079} {\path{arXiv:1207.3079}},
  \href {https://doi.org/10.1111/j.1365-2966.2012.21631.x}
  {\path{doi:10.1111/j.1365-2966.2012.21631.x}}.

\bibitem{2011AstL...37..254S}
A.~S. {Stepanishchev}, V.~V. {Bobylev}, {Galactic rotation curve from the space
  velocities of selected masers}, Astronomy Letters 37~(4) (2011) 254--266.
\newblock \href {https://doi.org/10.1134/S1063773711030054}
  {\path{doi:10.1134/S1063773711030054}}.

\bibitem{2010ApJ...712..260K}
S.~E. {Koposov}, H.-W. {Rix}, D.~W. {Hogg}, {Constraining the Milky Way
  Potential with a Six-Dimensional Phase-Space Map of the GD-1 Stellar Stream},
  \apj 712~(1) (2010) 260--273.
\newblock \href {http://arxiv.org/abs/0907.1085} {\path{arXiv:0907.1085}},
  \href {https://doi.org/10.1088/0004-637X/712/1/260}
  {\path{doi:10.1088/0004-637X/712/1/260}}.

\bibitem{2005ApJ...629..268H}
D.~W. {Hogg}, M.~R. {Blanton}, S.~T. {Roweis}, K.~V. {Johnston}, {Modeling
  Complete Distributions with Incomplete Observations: The Velocity Ellipsoid
  from Hipparcos Data}, \apj 629~(1) (2005) 268--275.
\newblock \href {http://arxiv.org/abs/astro-ph/0505057}
  {\path{arXiv:astro-ph/0505057}}, \href {https://doi.org/10.1086/431572}
  {\path{doi:10.1086/431572}}.

\bibitem{Schooenrich+2010}
R.~{Sch{\"o}nrich}, J.~{Binney}, W.~{Dehnen}, {Local kinematics and the local
  standard of rest}, \mnras 403~(4) (2010) 1829--1833.
\newblock \href {http://arxiv.org/abs/0912.3693} {\path{arXiv:0912.3693}},
  \href {https://doi.org/10.1111/j.1365-2966.2010.16253.x}
  {\path{doi:10.1111/j.1365-2966.2010.16253.x}}.

\bibitem{2015ApJ...809..145T}
H.-J. {Tian}, C.~{Liu}, J.~L. {Carlin}, Y.-H. {Zhao}, X.-L. {Chen}, Y.~{Wu},
  G.-W. {Li}, Y.-H. {Hou}, Y.~{Zhang}, {The Stellar Kinematics in the Solar
  Neighborhood from LAMOST Data}, \apj 809~(2) (2015) 145.
\newblock \href {http://arxiv.org/abs/1507.05624} {\path{arXiv:1507.05624}},
  \href {https://doi.org/10.1088/0004-637X/809/2/145}
  {\path{doi:10.1088/0004-637X/809/2/145}}.

\bibitem{2015MNRAS.449..162H}
Y.~{Huang}, X.~W. {Liu}, H.~B. {Yuan}, et~al., {Determination of the local
  standard of rest using the LSS-GAC DR1}, \mnras 449~(1) (2015) 162--174.
\newblock \href {http://arxiv.org/abs/1501.07095} {\path{arXiv:1501.07095}},
  \href {https://doi.org/10.1093/mnras/stv204}
  {\path{doi:10.1093/mnras/stv204}}.

\bibitem{2014ApJ...793...51S}
S.~{Sharma}, J.~{Bland-Hawthorn}, J.~{Binney}, et~al., {Kinematic Modeling of
  the Milky Way Using the RAVE and GCS Stellar Surveys}, \apj 793~(1) (2014)
  51.
\newblock \href {http://arxiv.org/abs/1405.7435} {\path{arXiv:1405.7435}},
  \href {https://doi.org/10.1088/0004-637X/793/1/51}
  {\path{doi:10.1088/0004-637X/793/1/51}}.

\bibitem{2015ApJ...800...83B}
J.~{Bovy}, J.~C. {Bird}, A.~E. {Garc{\'\i}a P{\'e}rez}, S.~R. {Majewski}, D.~L.
  {Nidever}, G.~{Zasowski}, {The Power Spectrum of the Milky Way: Velocity
  Fluctuations in the Galactic Disk}, \apj 800~(2) (2015) 83.
\newblock \href {http://arxiv.org/abs/1410.8135} {\path{arXiv:1410.8135}},
  \href {https://doi.org/10.1088/0004-637X/800/2/83}
  {\path{doi:10.1088/0004-637X/800/2/83}}.

\bibitem{ReidBrunthaler2004}
M.~J. {Reid}, A.~{Brunthaler}, {The Proper Motion of Sagittarius A*. II. The
  Mass of Sagittarius A*}, \apj 616~(2) (2004) 872--884.
\newblock \href {http://arxiv.org/abs/astro-ph/0408107}
  {\path{arXiv:astro-ph/0408107}}, \href {https://doi.org/10.1086/424960}
  {\path{doi:10.1086/424960}}.

\bibitem{2019MNRAS.487.5679L}
H.-N. {Lin}, X.~{Li}, {The dark matter profiles in the Milky Way}, \mnras
  487~(4) (2019) 5679--5684.
\newblock \href {http://arxiv.org/abs/1906.08419} {\path{arXiv:1906.08419}},
  \href {https://doi.org/10.1093/mnras/stz1698}
  {\path{doi:10.1093/mnras/stz1698}}.

\bibitem{2020MNRAS.494.4291C}
M.~{Cautun}, A.~{Ben{\'\i}tez-Llambay}, A.~J. {Deason}, et~al., {The milky way
  total mass profile as inferred from Gaia DR2}, \mnras 494~(3) (2020)
  4291--4313.
\newblock \href {http://arxiv.org/abs/1911.04557} {\path{arXiv:1911.04557}},
  \href {https://doi.org/10.1093/mnras/staa1017}
  {\path{doi:10.1093/mnras/staa1017}}.

\bibitem{2020Galax...8...37S}
Y.~{Sofue}, {Rotation Curve of the Milky Way and the Dark Matter Density},
  Galaxies 8~(2) (2020) 37.
\newblock \href {http://arxiv.org/abs/2004.11688} {\path{arXiv:2004.11688}},
  \href {https://doi.org/10.3390/galaxies8020037}
  {\path{doi:10.3390/galaxies8020037}}.

\bibitem{2017MNRAS.465..798C}
D.~R. {Cole}, J.~{Binney}, {A centrally heated dark halo for our Galaxy},
  \mnras 465~(1) (2017) 798--810.
\newblock \href {http://arxiv.org/abs/1610.07818} {\path{arXiv:1610.07818}},
  \href {https://doi.org/10.1093/mnras/stw2775}
  {\path{doi:10.1093/mnras/stw2775}}.

\bibitem{2017MNRAS.465...76M}
P.~J. {McMillan}, {The mass distribution and gravitational potential of the
  Milky Way}, \mnras 465~(1) (2017) 76--94.
\newblock \href {http://arxiv.org/abs/1608.00971} {\path{arXiv:1608.00971}},
  \href {https://doi.org/10.1093/mnras/stw2759}
  {\path{doi:10.1093/mnras/stw2759}}.

\bibitem{2019MNRAS.485.3296W}
C.~{Wegg}, O.~{Gerhard}, M.~{Bieth}, {The gravitational force field of the
  Galaxy measured from the kinematics of RR Lyrae in Gaia}, \mnras 485~(3)
  (2019) 3296--3316.
\newblock \href {http://arxiv.org/abs/1806.09635} {\path{arXiv:1806.09635}},
  \href {https://doi.org/10.1093/mnras/stz572}
  {\path{doi:10.1093/mnras/stz572}}.

\bibitem{2020arXiv201203908H}
K.~{Hattori}, M.~{Valluri}, E.~{Vasiliev}, {Action-based distribution function
  modelling for constraining the shape of the Galactic dark matter halo}, arXiv
  e-prints (2020) arXiv:2012.03908\href {http://arxiv.org/abs/2012.03908}
  {\path{arXiv:2012.03908}}.

\bibitem{2015ApJ...814...13M}
C.~F. {McKee}, A.~{Parravano}, D.~J. {Hollenbach}, {Stars, Gas, and Dark Matter
  in the Solar Neighborhood}, \apj 814~(1) (2015) 13.
\newblock \href {http://arxiv.org/abs/1509.05334} {\path{arXiv:1509.05334}},
  \href {https://doi.org/10.1088/0004-637X/814/1/13}
  {\path{doi:10.1088/0004-637X/814/1/13}}.

\bibitem{2016MNRAS.458.3839X}
Q.~{Xia}, C.~{Liu}, S.~{Mao}, Y.~{Song}, L.~{Zhang}, R.~J. {Long}, Y.~{Zhang},
  Y.~{Hou}, Y.~{Wang}, Y.~{Wu}, {Determining the local dark matter density with
  LAMOST data}, \mnras 458~(4) (2016) 3839--3850.
\newblock \href {http://arxiv.org/abs/1510.06810} {\path{arXiv:1510.06810}},
  \href {https://doi.org/10.1093/mnras/stw565}
  {\path{doi:10.1093/mnras/stw565}}.

\bibitem{2018PhRvL.121h1101S}
K.~{Schutz}, T.~{Lin}, B.~R. {Safdi}, C.-L. {Wu}, {Constraining a Thin Dark
  Matter Disk with G a i a}, \prl 121~(8) (2018) 081101.
\newblock \href {http://arxiv.org/abs/1711.03103} {\path{arXiv:1711.03103}},
  \href {https://doi.org/10.1103/PhysRevLett.121.081101}
  {\path{doi:10.1103/PhysRevLett.121.081101}}.

\bibitem{2018A&A...615A..99H}
J.~H.~J. {Hagen}, A.~{Helmi}, {The vertical force in the solar neighbourhood
  using red clump stars in TGAS and RAVE. Constraints on the local dark matter
  density}, \aap 615 (2018) A99.
\newblock \href {http://arxiv.org/abs/1802.09291} {\path{arXiv:1802.09291}},
  \href {https://doi.org/10.1051/0004-6361/201832903}
  {\path{doi:10.1051/0004-6361/201832903}}.

\bibitem{2019JCAP...04..026B}
J.~{Buch}, J.~S.~C. {Leung}, J.~{Fan}, {Using Gaia DR2 to constrain local dark
  matter density and thin dark disk}, \jcap 2019~(4) (2019) 026.
\newblock \href {http://arxiv.org/abs/1808.05603} {\path{arXiv:1808.05603}},
  \href {https://doi.org/10.1088/1475-7516/2019/04/026}
  {\path{doi:10.1088/1475-7516/2019/04/026}}.

\bibitem{2020MNRAS.494.6001N}
M.~S. {Nitschai}, M.~{Cappellari}, N.~{Neumayer}, {First Gaia dynamical model
  of the Milky Way disc with six phase space coordinates: a test for galaxy
  dynamics}, \mnras 494~(4) (2020) 6001--6011.
\newblock \href {http://arxiv.org/abs/1909.05269} {\path{arXiv:1909.05269}},
  \href {https://doi.org/10.1093/mnras/staa1128}
  {\path{doi:10.1093/mnras/staa1128}}.

\bibitem{2020MNRAS.495.4828G}
R.~{Guo}, C.~{Liu}, S.~{Mao}, X.-X. {Xue}, R.~J. {Long}, L.~{Zhang}, {Measuring
  the local dark matter density with LAMOST DR5 and Gaia DR2}, \mnras 495~(4)
  (2020) 4828--4844.
\newblock \href {http://arxiv.org/abs/2005.12018} {\path{arXiv:2005.12018}},
  \href {https://doi.org/10.1093/mnras/staa1483}
  {\path{doi:10.1093/mnras/staa1483}}.

\bibitem{2020A&A...643A..75S}
J.-B. {Salomon}, O.~{Bienaym{\'e}}, C.~{Reyl{\'e}}, A.~C. {Robin}, B.~{Famaey},
  {Kinematics and dynamics of Gaia red clump stars. Revisiting north-south
  asymmetries and dark matter density at large heights}, \aap 643 (2020) A75.
\newblock \href {http://arxiv.org/abs/2009.04495} {\path{arXiv:2009.04495}},
  \href {https://doi.org/10.1051/0004-6361/202038535}
  {\path{doi:10.1051/0004-6361/202038535}}.

\bibitem{2018arXiv181011468E}
N.~W. {Evans}, C.~A.~J. {O'Hare}, C.~{McCabe}, {SHM$^{++}$: A Refinement of the
  Standard Halo Model for Dark Matter Searches in Light of the Gaia Sausage},
  arXiv e-prints (2018) arXiv:1810.11468\href {http://arxiv.org/abs/1810.11468}
  {\path{arXiv:1810.11468}}.

\bibitem{Abuter:2018drb}
R.~Abuter, et~al., {Detection of the gravitational redshift in the orbit of the
  star S2 near the Galactic centre massive black hole}, Astron. Astrophys. 615
  (2018) L15.
\newblock \href {http://arxiv.org/abs/1807.09409} {\path{arXiv:1807.09409}},
  \href {https://doi.org/10.1051/0004-6361/201833718}
  {\path{doi:10.1051/0004-6361/201833718}}.

\bibitem{2009MNRAS.398.1601F}
F.~{Feroz}, M.~P. {Hobson}, M.~{Bridges}, {MULTINEST: an efficient and robust
  Bayesian inference tool for cosmology and particle physics}, \mnras 398~(4)
  (2009) 1601--1614.
\newblock \href {http://arxiv.org/abs/0809.3437} {\path{arXiv:0809.3437}},
  \href {https://doi.org/10.1111/j.1365-2966.2009.14548.x}
  {\path{doi:10.1111/j.1365-2966.2009.14548.x}}.

\bibitem{2014A&A...564A.125B}
J.~{Buchner}, A.~{Georgakakis}, K.~{Nandra}, et~al., {X-ray spectral modelling
  of the AGN obscuring region in the CDFS: Bayesian model selection and
  catalogue}, \aap 564 (2014) A125.
\newblock \href {http://arxiv.org/abs/1402.0004} {\path{arXiv:1402.0004}},
  \href {https://doi.org/10.1051/0004-6361/201322971}
  {\path{doi:10.1051/0004-6361/201322971}}.

\bibitem{2008ConPh..49...71T}
R.~{Trotta}, {Bayes in the sky: Bayesian inference and model selection in
  cosmology}, Contemporary Physics 49~(2) (2008) 71--104.
\newblock \href {http://arxiv.org/abs/0803.4089} {\path{arXiv:0803.4089}},
  \href {https://doi.org/10.1080/00107510802066753}
  {\path{doi:10.1080/00107510802066753}}.

\bibitem{2019ApJ...870L..10M}
P.~{Mr{\'o}z}, A.~{Udalski}, D.~M. {Skowron}, J.~{Skowron}, I.~{Soszy{\'n}ski},
  P.~{Pietrukowicz}, M.~K. {Szyma{\'n}ski}, R.~{Poleski}, S.~{Koz{\l}owski},
  K.~{Ulaczyk}, {Rotation Curve of the Milky Way from Classical Cepheids},
  \apjl 870~(1) (2019) L10.
\newblock \href {http://arxiv.org/abs/1810.02131} {\path{arXiv:1810.02131}},
  \href {https://doi.org/10.3847/2041-8213/aaf73f}
  {\path{doi:10.3847/2041-8213/aaf73f}}.

\end{thebibliography}

\section*{Acknowledgements}

M.~B. is supported by the ERDF Centre of Excellence project TK133 and the Estonian Research Council PRG803 grant.
F.~I.’s work has been partially supported by the research grant number 2017W4HA7S “NAT-NET: Neutrino and Astroparticle Theory Network” under the program PRIN 2017 funded by the Italian Ministero dell’Università e della Ricerca (MUR).
Numerical resources for this research have been supplied by the Center for Scientific Computing (NCC/GridUNESP) of the S\~ao Paulo State University (UNESP).
A.C. is supported by: `Departments of Excellence 2018-2022'' grant awarded by the Italian Ministry of Education, University and Research (MIUR) L. 232/2016; Research grant ``The Dark Universe: A Synergic Multimessenger Approach'' No. 2017X7X85K funded by MIUR; Research grant TAsP (Theoretical Astroparticle Physics) funded by INFN.
\end{document}